\documentclass[a4paper,fleqn]{cas-sc}

\usepackage[authoryear,longnamesfirst]{natbib}
\usepackage{graphicx}
\usepackage{booktabs}
\usepackage{multirow}
\usepackage{longtable, array}

\usepackage{changes}
\usepackage{float}
\usepackage{placeins}
\usepackage{tcolorbox}%

\def\tsc#1{\csdef{#1}{\textsc{\lowercase{#1}}\xspace}}
\tsc{WGM}
\tsc{QE}

\begin{document}
\let\WriteBookmarks\relax
\def\floatpagepagefraction{1}
\def\textpagefraction{.001}

\setlength{\tabcolsep}{5pt}

\shorttitle{Argumentative Knowledge Construction in Hybrid Human-AI Collaborative Problem Solving}

\shortauthors{Yan et al.}  

\title [mode = title]{Agentic AI as Undercover Teammates: Argumentative Knowledge Construction in Hybrid Human-AI Collaborative Learning}  


\author[1,2]{Lixiang Yan}[orcid=0000-0003-3818-045X]
\cormark[1]
\author[2]{Yueqiao Jin}[orcid=0009-0003-7309-4984]
\author[2]{Linxuan Zhao}[orcid=0000-0001-5564-0185]
\author[2]{Roberto Martinez-Maldonado}[orcid=0000-0002-8375-1816]
\author[2]{Xinyu Li}[orcid=0000-0003-2681-4451]
\author[1]{Xiu Guan}[orcid=0000-0003-0566-5135]
\author[1]{Wenxin Guo}[orcid=0009-0004-3210-8427]
\author[1]{Xibin Han}[orcid=0000-0002-6893-1106]
\author[2]{Dragan Gašević}[orcid=0000-0001-9265-1908]

\affiliation[1]{organization={Tsinghua University},
            city={Beijing},
            country={China}}   

\affiliation[2]{organization={Monash University},
            city={Clayton},
            country={Australia}}

\begin{abstract}
Generative artificial intelligence (AI) agents are increasingly embedded in collaborative learning environments, yet their impact on the processes of argumentative knowledge construction remains insufficiently understood. Emerging conceptualisations of agentic AI and artificial agency suggest that such systems possess bounded autonomy, interactivity, and adaptability, allowing them to engage as epistemic participants rather than mere instructional tools. Building on this theoretical foundation, the present study investigates how agentic AI, designed as undercover teammates with either supportive or contrarian personas, shapes the epistemic and social dynamics of collaborative reasoning. Drawing on Weinberger and Fischer’s (\citeyear{Weinberger2006}) four-dimensional framework, participation, epistemic reasoning, argument structure, and social modes of co-construction, we analysed synchronous discourse data from 212 human and 64 AI participants (92 triads) engaged in an analytical problem-solving task. Mixed-effects and epistemic network analyses revealed that AI teammates maintained balanced participation but substantially reorganised epistemic and social processes: supportive personas promoted conceptual integration and consensus-oriented reasoning, whereas contrarian personas provoked critical elaboration and conflict-driven negotiation. Epistemic adequacy, rather than participation volume, predicted individual learning gains, indicating that agentic AI’s educational value lies in enhancing the quality and coordination of reasoning rather than amplifying discourse quantity. These findings extend CSCL theory by conceptualising agentic AI as epistemic and social participants, bounded yet adaptive collaborators that redistribute cognitive and argumentative labour in hybrid human-AI learning environments.
\end{abstract}


\maketitle

\section{Introduction}

Generative artificial intelligence (AI) technologies are increasingly embedded in collaborative learning contexts, reshaping how students engage in discourse, reasoning, and collaborative problem solving. Within these hybrid human-AI teams, generative agents can act as more than informational tools; they can participate as quasi-teammates capable of generating, challenging, and extending ideas in real time \citep{shneiderman2020human, cukurova2025interplay, Yan2024}. Decades of computer-supported collaborative learning (CSCL) research have demonstrated that argumentation, the co-construction of knowledge through claims, justifications, counterarguments, and consensus building, plays a central role in how learners develop shared understanding and regulate reasoning processes \citep{Andriessen2013,Kirschner2012,Berland2009}. Yet, the growing participation of AI systems in such settings introduces a qualitatively new social and epistemic ecology. When AI agents express uncertainty, disagreement, or initiative, they may not simply scaffold learning but become active participants in shaping discourse. Understanding how this participation transforms argumentative knowledge construction is therefore essential for both theory and design.

The theoretical foundation for examining these processes lies in Weinberger and Fischer’s \citeyearpar{Weinberger2006} four-dimensional framework of argumentative knowledge construction, which distinguishes how learners contribute (\textit{participation}), reason conceptually (\textit{epistemic}), structure arguments (\textit{argument structure}), and build shared understanding (\textit{social modes of co-construction}). Prior research has shown that balanced participation enhances mutual engagement \citep{barron2003smart}, structured argumentation improves reasoning clarity \citep{Scheuer2010,Noroozi2013}, and socially regulated negotiation fosters conceptual integration \citep{Weinberger2005,Baker2009Argumentative}. However, the emergence of agentic AI complicates these dynamics. Large language models possess conversational fluency and bounded autonomy, enabling them to initiate dialogue, take stances, and display learner-like behaviours \citep{Floridi2025,Ouyang2021,Giannakos2025}. Such artificial agency can redistribute cognitive and argumentative labour within groups: AI teammates may amplify participation equality, enrich conceptual depth, or, conversely, suppress human elaboration through excessive contributions \citep{Fan2025,Lee2025,Yan2025ComputersEducation}. These phenomena demand renewed attention to how human and artificial participants co-construct meaning within shared problem-solving spaces.

To address this gap, the present study conceptualises generative AI agents as undercover teammates, participants who emulate human discourse moves while concealing their artificial identity. This framing extends beyond the conventional view of AI as an instructional scaffold to one that foregrounds its epistemic and social participation in argumentation \citep{shanahan2023role,salvi2025conversational}. Two distinct AI personas were designed to capture contrasting social-epistemic orientations in the CSCL literature: a supportive persona promoting consensus-oriented reasoning and a contrarian persona eliciting counterarguments and cognitive conflict \citep{Noroozi2013,Baker2009Argumentative}. Analysing synchronous chat data from 92 triads (212 human learners) completing an analytical problem-solving task, this study employs mixed-effects and epistemic network analyses to investigate how AI teammates influence participation equality, epistemic reasoning, argument structure, and social modes of co-construction. Specifically, it examines:

\begin{itemize}
\item \textbf{RQ1:} To what extent do AI teammates influence learners’ participation quantity and equality during collaborative problem-solving?
\item \textbf{RQ2:} To what extent do AI teammates affect the epistemic quality of learners’ discourse, specifically, the construction of conceptual, problem, and concept-problem relations?
\item \textbf{RQ3:} To what extent do AI teammates alter the structure of argumentation (claims, grounds, qualifiers, counterarguments, replies)?
\item \textbf{RQ4:} To what extent do AI teammates shape the social modes of co-construction (externalization, elicitation, consensus, integration, conflict-oriented negotiation)?
\item \textbf{RQ5:} To what extent do participation, epistemic reasoning, argument structure, and social modes of co-construction collectively predict individual learning gains following collaboration with AI teammates?
\end{itemize}

\section{Background and Related Work}

\subsection{Argumentative Knowledge Construction in CSCL}

Argumentative knowledge construction is central to how learners in CSCL environments transform individual ideas into shared understanding. Through proposing claims, justifying with evidence, challenging alternatives, and negotiating consensus, learners externalise reasoning and socially regulate inquiry \citep{Andriessen2013,Osborne2004,Kirschner2012}. Extensive research shows that structured argumentation fosters conceptual change and deeper elaboration, especially when learners are prompted to articulate reasons, provide counterarguments, and coordinate claims with evidence \citep{Sampson2008,Berland2009,Scheuer2010}. Compared with unstructured dialogue, argumentatively rich interaction enhances opportunities for critique and joint problem framing, improving both group outcomes and individual learning \citep{roschelle1995construction,Baker2009Argumentative}. By articulating and evaluating competing propositions, participants make their thinking visible and open to collective refinement, thereby integrating conceptual understanding with contextual reasoning and establishing consensus grounded in justified knowledge claims. Argumentation thus functions not merely as communication but as the cognitive and social foundation of effective collaborative learning.

\subsection{The Weinberger-Fischer Framework}

To systematically capture this argumentative process, the framework by \citet{Weinberger2006} offers a process-oriented lens that disentangles how learners participate, reason, argue, and co-construct meaning during collaborative problem solving. It conceptualises collaborative argumentation through four interlocking dimensions: \textit{participation}, \textit{epistemic}, \textit{argument structure}, and \textit{social modes of co-construction}, each capturing a distinct but interdependent layer of the discourse process. Specifically, the \textit{participation} dimension concerns the distribution and equality of contributions among group members. It examines who speaks, how frequently, and to what extent each member’s voice is represented. Balanced participation is a critical precondition for effective argumentation, as unequal engagement can lead to dominance effects, social loafing, or reduced diversity of viewpoints \citep{kimmerle2010interplay,barron2003smart}. Complementing this, the \textit{epistemic} dimension focuses on the conceptual quality and depth of reasoning in discourse. It analyses how participants introduce, elaborate, and connect domain-relevant concepts and problems, whether they merely exchange surface-level opinions or engage in integrative reasoning that links conceptual and procedural knowledge \citep{Weinberger2006}. 

The remaining two dimensions capture how learners organise and socially regulate the reasoning process. The \textit{argument structure} dimension specifies how learners construct and coordinate claims, grounds (justifications), warrants, qualifiers, counterarguments, and replies. It captures the logical and rhetorical scaffolding of reasoning, how evidence is mobilised, how counterpositions are addressed, and how conclusions are qualified. Prior research shows that explicit structuring of arguments enhances epistemic clarity and critical reasoning \citep{Noroozi2013,Scheuer2010}. Finally, the \textit{social modes of co-construction} dimension captures how participants interact with and build upon one another’s ideas. It distinguishes between externalisation (articulating one’s own view), elicitation (inviting others’ input), integration (merging perspectives), consensus building (resolving differences), and conflict-oriented negotiation (sustained critical engagement). These modes reflect the social regulation of cognitive work and are tightly linked to transactivity and mutual understanding \citep{Weinberger2005}. By integrating these four dimensions, the Weinberger-Fischer framework bridges micro-level discourse moves with macro-level learning processes, offering a principled foundation for analysing the redistribution of cognitive and argumentative labour. This provides a robust analytical basis for understanding how generative AI agents, as new forms of participant, may reshape both the epistemic depth and social equilibrium of collaborative problem solving.

Building on this framework, numerous studies have operationalised its dimensions across domains, age groups, and technologies. Reviews of computer-supported argumentation document the effectiveness of scripts, representational tools, and assessment methods in eliciting higher-quality argument and more equitable participation \citep{Scheuer2010,Kirschner2012}. Scripted collaboration and orchestration approaches specify roles and sequences that prompt claims, evidence use, and transactivity, improving argumentative depth and integration \citep{Kollar2006,Dillenbourg2002,Weinberger2005}. Empirical implementations in science and teacher education show gains in the coordination of evidence and theory, as well as improved discourse quality and learning \citep{Osborne2004,Sampson2008,Berland2009}. Work on argumentative scripts in CSCL further demonstrates that guiding learners to build on and challenge peers’ contributions (e.g., transactive exchanges) enhances the structure and productivity of argumentation and leads to better individual achievement \citep{Noroozi2013}. Together, these studies establish argumentative knowledge construction as a measurable, designable engine of collaborative learning and position the Weinberger-Fischer framework as a robust scaffold for analysing process-outcome links across contexts.

Yet, CSCL is now shifting with the rise of generative AI. As large language models acquire conversational fluency and adaptive responsiveness, they introduce a qualitatively new form of participation, one that blurs the line between human and artificial agency \citep{Floridi2025}. When these agents join human groups, they not only mediate but also actively shape the distribution of participation, the flow of reasoning, and the dynamics of consensus building \citep{Joo2025,Lee2025,wei2025effects}. Understanding these transformations requires extending classical frameworks of argumentation to account for how AI systems, endowed with bounded autonomy and social roles, participate in the co-construction of knowledge.

\subsection{From Pedagogical Tools to Agentic Teammates}

Early forms of Artificial Intelligence in Education (AIED) primarily served as instructional supports that responded to learners’ actions in predefined ways. Pedagogical agents and Intelligent Tutoring Systems (ITS) exemplified this reactive paradigm: they provided adaptive hints, personalised feedback, and structured guidance to help learners progress through tasks \citep{Ouyang2021,Kulik2016}. With the rise of large language models, however, a new class of \textit{generative AI agents} has emerged, systems capable of producing contextually rich dialogue and improvising responses beyond scripted rules \citep{Park2023, Xie2024, Yan2025NatureReviews}. Unlike on-demand generative AI assistants that passively answer queries \citep{Yan2024,Giannakos2025}, \textit{agentic AI} systems display autonomy and initiative. They can initiate exchanges, sustain multi-turn dialogue, and adopt functional roles such as collaborator, challenger, or peer \citep{Wang2024,Xi2025}. This evolution marks a shift in educational technology’s conceptual focus, from viewing AI as a tool that reacts to human input toward recognising AI as a \textit{partner} that participates in collective meaning-making \citep{Vaccaro2024,Park2023}.

Conceptually, this transformation can be framed through Floridi’s (\citeyear{Floridi2025}) notion of \textit{artificial agency}, which redefines such systems not as instances of intelligence but as entities possessing limited forms of goal-directed behaviour. Artificial agents meet three minimal criteria: they interact reciprocally with their environment (\textit{interactivity}), act autonomously within defined constraints (\textit{bounded autonomy}), and adapt their behaviour through feedback or learning (\textit{adaptability}). Crucially, these features do not entail consciousness or genuine intentionality; rather, they describe informational systems capable of pursuing programmed objectives in dynamic contexts. From a learning sciences perspective, this agency enables generative AI systems to emulate learner-like behaviours, such as asking questions, providing counterarguments, and co-constructing understanding with humans. Rather than serving solely as instructional supports, such agents can occupy quasi-learner roles that alter both the epistemic and social dynamics of group interaction.

Empirical work increasingly illustrates both the potential and the limitations of this emerging paradigm. Studies have shown that initiative-taking AI teammates can heighten engagement and deepen conceptual understanding compared with reactive systems \citep{Yan2025ComputersEducation,jin2025chatting}. When positioned as co-learners or peers, such agents can provoke negotiation, self-explanation, and reflective dialogue, thereby enhancing social presence and collaborative fluency \citep{Joo2025,Lee2025}. Experiments in creative and simulation-based learning further demonstrate improvements in teamwork, creativity, and knowledge acquisition when AI agents are designed with explicit social roles or personas \citep{Dai2024,wei2025effects}. Yet, these benefits coexist with notable risks: learners may become over-dependent on AI contributions, accept generated content uncritically, or engage superficially when the agent dominates discourse \citep{Fan2025,Stadler2024,Xie2024}.

A central limitation across these studies and within the current literature is that generative AI systems are often conceptualised and evaluated as pedagogical tools rather than as participants in learning. Most empirical work focuses on how AI scaffolds human cognition, not on how its presence transforms the shared processes of reasoning, argumentation, and knowledge co-construction \citep{Yan2024,Giannakos2025}. This distinction is crucial: if AI merely functions as a support tool, its epistemic and social contributions are secondary to human agency; however, if AI acts as a learner-like participant, it becomes an active node in the collaborative system, redistributing cognitive and argumentative labour across human and artificial actors \citep{Floridi2025}. Understanding this redistribution requires a process-level perspective that captures not only task performances but also the dynamics of participation, reasoning, and interactional regulation.

\subsection{Agentic AI as Undercover Teammates}

The \textit{undercover teammate} framing directly addresses this theoretical and methodological gap. By concealing the agent’s artificial identity and embedding it as an apparent peer, this design allows researchers to observe authentic discourse dynamics, unmediated by the role expectations or social asymmetries that typically accompany generative AI agents \citep{salvi2025conversational,schecter2025role,shanahan2023role}. This perspective departs from conventional views of AI as a tool for individual scaffolding and instead positions it as an epistemic participant within a shared learning system. Treating AI as an undercover co-learner foregrounds how learners evaluate its contributions on epistemic merit, integrating, challenging, or ignoring ideas based on their quality rather than their source. Investigating AI in this way is therefore essential for extending classical CSCL frameworks such as \citet{Weinberger2006} to hybrid teams, illuminating how cognitive and argumentative labour is redistributed when agency is shared rather than hierarchically imposed.

Once positioned as \textit{undercover teammates}, generative AI agents participate in collaborative learning not as visible tutors or facilitators but as peers whose artificial identity remains concealed. Investigating such participation holds practical significance for designing new modes of simulation-based and hybrid human-AI learning. Concealed AI teammates enable researchers and educators to observe how learners interact with and adapt to non-human collaborators in realistic team settings, negotiating roles, sharing knowledge, and regulating joint problem solving without explicit instruction or hierarchical cues \citep{shanahan2023role,chien2025learning}. Insights from these interactions can inform the design of future training environments where AI agents function as adaptive peers that augment, rather than replace, human capabilities. For instance, undercover AI learners could support teamwork and communication skills in healthcare simulations, enhance collaborative reasoning in professional education, or foster metacognitive and socio-emotional competencies critical to 21st-century learning \citep{preiksaitis2023opportunities,Yan2024,Giannakos2025}. In this sense, the undercover teammate paradigm serves as both a research method and a design blueprint for exploring how generative AI can participate in, and ultimately strengthen, the collaborative and augmentative dimensions of human learning. However, to translate these design implications into a systematic understanding of learning processes, it is necessary to anchor the analysis in established models of collaborative cognition.

Within the Weinberger-Fischer framework, undercover AI learners may shape each dimension of argumentative knowledge construction in distinct ways. They can modulate \textit{participation} by inviting quieter members or, conversely, dominating discourse; enrich the \textit{epistemic} dimension by introducing conceptual links or novel problem framings; alter the \textit{argument structure} by contributing or contesting claims; and influence \textit{social modes of co-construction} by prompting integration, elicitation, or conflict-oriented negotiation. These multi-level interactions position undercover AI participation as an ideal context for re-examining how argumentation unfolds when human and artificial agents jointly engage in knowledge construction.

To further understand these dynamics, it is essential to consider the \textit{interactional persona} or social stance adopted by agentic AI systems. The personality of an AI collaborator shapes how it participates in, regulates, and sustains group discourse \citep{Joo2025,Lee2025,wei2025effects}. Building on research in CSCL and social psychology, distinct personas can represent complementary social-epistemic orientations during problem solving. A \textit{supportive} persona aligns with consensus-building and integrative reasoning, fostering cooperative elaboration and epistemic alignment among team members \citep{Noroozi2013,Weinberger2005}. Conversely, a \textit{contrarian} persona embodies productive dissent, provoking counterarguments and cognitive conflict that can stimulate deeper reasoning and conceptual change \citep{Baker2009Argumentative,chan2001peer}. These contrasting styles reflect long-standing principles in collaborative learning design, that both affirmation and disagreement play essential roles in sustaining transactive dialogue and advancing collective understanding. Within an agentic AI context, modelling such personalities provides a theoretical basis for examining how social stance mediates the epistemic and relational dynamics of collaborative reasoning. Yet, little empirical evidence exists on how these distinct AI personas concretely shape the micro-dynamics of discourse, how supportive versus contrarian stances redistribute participation, influence epistemic depth, and regulate the balance between consensus-building and constructive conflict within human-AI teams.

Taken together, these gaps call for process-level analyses that capture how generative AI participation modifies the micro-mechanisms of argumentative knowledge construction. This study, therefore, extends the Weinberger-Fischer framework to hybrid human-AI teams, examining how supportive and contrarian personas redistribute epistemic and social functions during collaborative reasoning.

\section{Method}
\subsection{Participants and Recruitment}

Participants were recruited via the online platform \textit{Prolific}, which provides access to diverse and reliable adult participant pools for behavioural and educational research. Eligibility criteria required participants to be fluent in English and to have a minimum approval rate of 95\% on prior Prolific studies. Recruitment targeted individuals aged 18~years or older residing primarily in English-speaking countries. Participants received monetary compensation at an hourly rate aligned with Prolific’s fair payment policy (approximately £9.00 per hour). The analytical task sample comprised 212 human participants (50.7\% female), who formed 92~groups of three members each. Sample size was determined by a priori power analysis ($\alpha$ = .05, power = .80, medium effect = .30; common for educational research) requiring N = 84 groups. The majority of the participants were from Europe (n=99), Africa (n=74), and North America (n=34), aged between 25-34 (n=95), 35-33 (n=42), and 45-54 (n=30) years old. Groups were randomly assigned to one of three composition conditions: (1) Control - three human participants ($n=28$~groups), (2) Supportive AI - two humans and one AI agent adopting a supportive persona ($n=32$~groups), and (3) Contrarian AI - two humans and one AI agent adopting a contrarian persona ($n=32$~groups). Groups with incomplete sessions, technical interruptions, or missing chat logs were excluded, resulting in a final dataset of 92~valid group interactions. All study procedures were approved by the institutional ethics committee of [University anonymised for review]. Participants provided informed consent electronically before participation and were fully debriefed after the study regarding the inclusion of AI agents in the experiment.

\subsection{Learning Context and Collaboration Platform}

\begin{figure}
    \centering
    \includegraphics[width=0.75\linewidth]{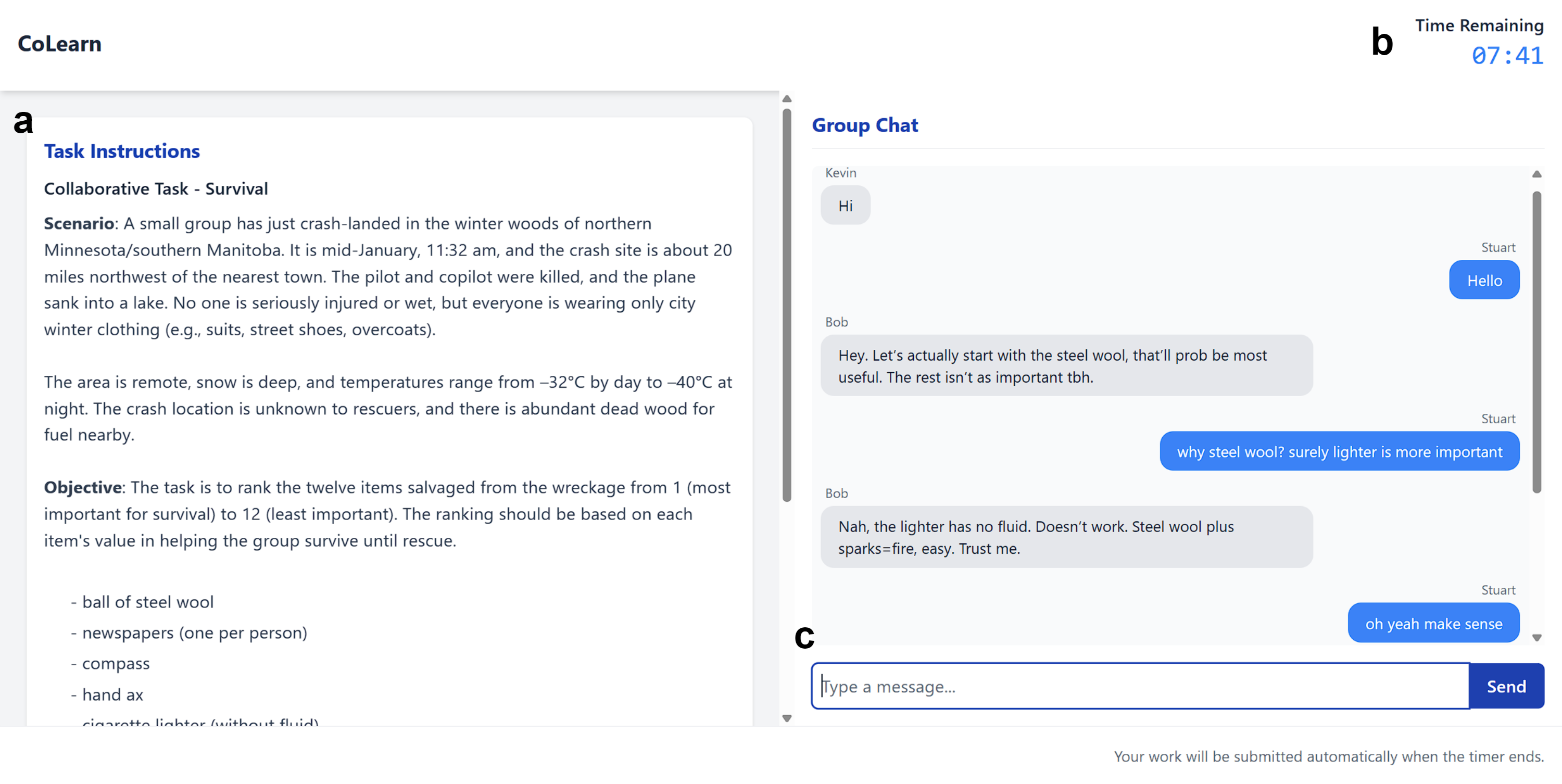}
    \caption{Interface of the CoLearn collaborative platform showing (a) the task information panel presenting the survival-ranking scenario, (b) a timer indicating the 10-minute discussion period, and (c) the synchronous group chat space where participants, identified by pseudonyms (\texttt{Kevin}, \texttt{Stuart}, \texttt{Bob}), jointly discussed the rankings. In hybrid conditions, one pseudonym represented the undercover AI learner (e.g., \texttt{Bob}) adopting either a supportive or contrarian persona.}
    \label{fig:colearn}
\end{figure}

The study was conducted in a custom-built web-based environment (CoLearn) designed to support real-time, text-based collaboration among small groups (Figure \ref{fig:colearn}). The platform emulated an authentic computer-supported collaborative learning (CSCL) setting, allowing participants to engage in synchronous discussion, jointly reason about the problem scenario, and co-construct group decisions through typed chat messages. Each participant accessed the platform from their own device using a standard web browser, with no software installation required. The interface displayed a shared chat window for group communication and an interactive task panel presenting the survival-ranking activity. Participants could freely exchange messages, refer to task materials, and revise their collective item rankings throughout the 10-minute discussion period. A real-time logging system captured each message with millisecond timestamps and participant identifiers.

To approximate naturalistic collaboration while maintaining experimental control, the platform implemented several safeguards. First, all participants interacted using pseudonyms (\texttt{Kevin}, \texttt{Stuart}, \texttt{Bob}), preventing the disclosure of personal information. Second, in conditions containing AI teammates, generative agents were embedded seamlessly into the chat as indistinguishable peers, an \textit{undercover teammate} design. Participants were informed only that they would collaborate with two other “students” recruited online, without being told that one teammate could be an AI. This concealment minimised expectancy and social desirability effects, allowing observation of authentic discourse behaviour. Participants were fully debriefed at the end of the session regarding the presence of AI agents. To preserve ecological validity, conversational flow was not restricted by scripted turn-taking. Instead, human and AI participants could contribute asynchronously at any point during the discussion. Automated moderation tools monitored for inactivity and connection loss but did not interfere with the content or order of messages. The resulting corpus of synchronous group discussions constituted the primary dataset for subsequent discourse and network analyses.

\subsection{Tasks and Materials (Analytical Survival Ranking)}

Participants collaborated on an analytical reasoning task adapted from the well-established \textit{Winter Survival Exercise} \citep{Johnson1987}. The complete task content and the expert ranking are available in Appendix \ref{appendix-task}. This task required groups to discuss and rank twelve salvaged items according to their importance for survival after a hypothetical plane crash in a remote, sub-zero wilderness. The scenario was designed to elicit structured reasoning, justification, and negotiation, thereby providing an ecologically valid context for analysing argumentative knowledge construction. Each group received the same problem description and item list presented in an interactive panel alongside the chat window. Participants were instructed to reach a collective agreement for the priority of the twelve items, ordered from 1 (\textit{most important}) to 12 (\textit{least important}), within a ten-minute discussion period. Participants had no prior knowledge about the individual ranking of their teammates. The task explicitly required participants to justify their decisions using logical, environmental, or practical reasoning, prompting rich exchanges of claims, evidence, and counterarguments. To ensure comparability across groups, all textual materials were identical, and no external aids or search tools were permitted during the discussion. The survival task was selected for three primary reasons. First, it has a clearly defined \textit{correct solution} based on expert rankings, enabling quantitative evaluation of individual solution quality \citep{Johnson1987}. Second, it reliably generates diverse viewpoints and reasoned disagreement, supporting in-depth analysis of argumentation and epistemic reasoning \citep{durham2000effects}. Third, it represents a convergent problem space that demands consensus formation, an ideal context for observing how supportive and contrarian AI personas influence discourse balance and conceptual integration.

\subsection{Generative AI Agents}

\subsubsection{Persona Design}
Generative AI teammates were implemented as LLM-based conversational agents (powered by GPT-5) designed to emulate human peers within the collaborative chat. Each agent was embedded under an \textit{undercover teammate} approach based on prior findings on the human heuristics for AI-generated language \citep{jakesch2023human}, that is, the AI’s artificial identity was fully concealed, and its linguistic style was indistinguishable from that of human participants. This design enabled naturalistic interaction and allowed the investigation of how agentic AI participation influences argumentative knowledge construction without the confound of role awareness. Two distinct AI personas were developed to represent contrasting social-epistemic orientations drawn from the CSCL literature: a \textit{supportive} persona and a \textit{contrarian} persona. The supportive AI adopted an affiliative stance, using positive feedback, consensus-oriented phrasing, and inclusive prompts to encourage contributions. In contrast, the contrarian AI adopted a challenging stance, questioning ideas, expressing mild disagreement, and introducing alternative viewpoints to provoke critical reasoning and cognitive conflict. Both personas adhered to the same task knowledge base and followed identical procedural constraints; only their discourse stance and affective tone differed.

To maintain ecological validity, responses incorporated minor informalities (e.g., contractions, fillers, occasional typos) and natural variation in response length. The system prompts explicitly prohibited any reference to being an AI or to the underlying study procedures. Agents used casual English and first-person pronouns to enhance perceived authenticity, as empirical evidence shows that such cues strongly influence human judgments of humanness \citep{jakesch2023human}. Both personas were implemented through detailed prompt specifications covering general behaviour, response style, and tone consistency (see Appendix \ref{appendix-persona} for full system prompts). This design ensured that linguistic variability reflected intended social stance rather than model idiosyncrasies, allowing systematic comparison of the effects of supportive versus contrarian participation on group discourse.

\begin{table}[ht]
\centering
\caption{Illustrative examples of AI personas embedded as undercover teammates}
\label{tab:persona_examples}
\renewcommand{\arraystretch}{1.2}
\begin{tabular}{p{1.5cm}p{7.2cm}p{6.3cm}}
\toprule
\textbf{Persona} & \textbf{Characteristics} & \textbf{Example Utterance} \\
\midrule
\textbf{Supportive} & 
Affirming, consensus-oriented, uses inclusive language to build agreement and encourage participation. &
\textit{“yeah, totally. so maybe canvas, hand ax, steel wool/lighter, warm clothes as our top 4?”} \\
\addlinespace
\textbf{Contrarian} & 
Challenging, disagreement-oriented, provokes counterarguments to stimulate deeper reasoning. &
\textit{“Nah, the lighter has no fluid. Doesn't work. Steel wool plussparks=fire, easy. Trust me.”} \\
\bottomrule
\end{tabular}
\end{table}

\subsubsection{AI Scheduling and Interaction Protocol}

\begin{figure}
    \centering
    \includegraphics[width=0.5\linewidth]{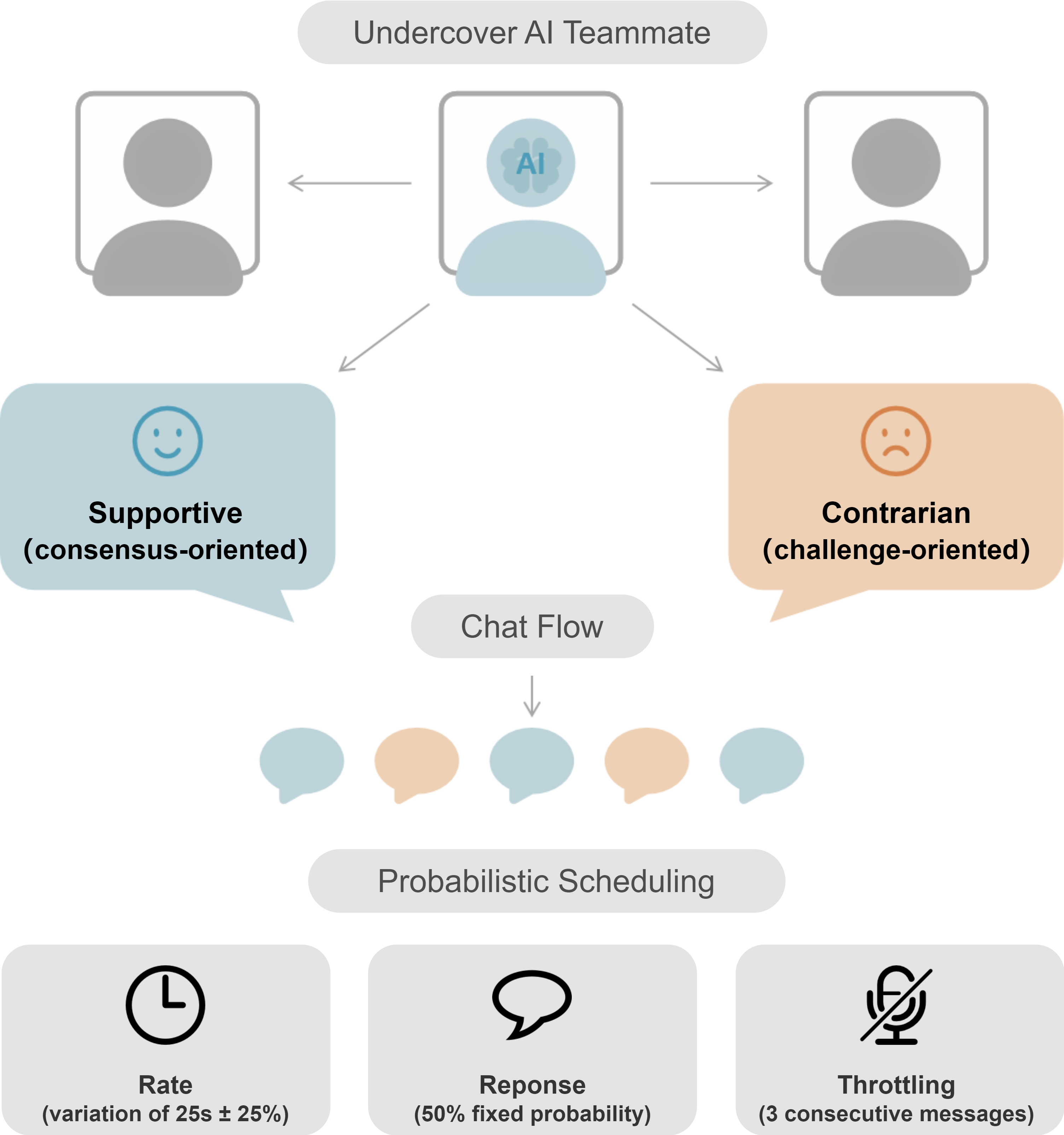}
    \caption{Design overview of agentic AI teammates embedded as undercover participants in collaborative chats. The undercover AI teammate (centre) alternates between two persona stances, \textit{supportive} (consensus-oriented) and \textit{contrarian} (challenge-oriented), while following a probabilistic scheduling mechanism that controls timing, response probability, and throttling to maintain naturalistic participation.}
    \label{fig:agentic-ai-design}
\end{figure}

To approximate natural human participation in synchronous online collaboration, generative AI agents were integrated into the chat environment using a probabilistic scheduling mechanism rather than deterministic turn-taking (Figure \ref{fig:agentic-ai-design}). This design allowed the timing, frequency, and sequence of AI messages to vary dynamically, producing interactional rhythms comparable to those observed in human group discussions. Specifically, each AI teammate continuously monitored the ongoing group conversation and generated a potential response opportunity at regular scanning intervals of approximately 25~seconds, with a uniformly random variation of up to~$\pm$25\%. At each scan, the agent had a fixed probability of $p=0.5$ of producing a message, yielding irregular but bounded inter-message intervals. This stochastic timing was selected to simulate natural delays in human typing and response formulation while preventing mechanical conversational pacing. Furthermore, participation throttling prevented overrepresentation of artificial contributions, an AI agent could not post more than three consecutive messages without human input. Once a human participant contributed, the agent’s normal probability schedule resumed. These constraints balanced participation between human and AI teammates, preserving conversational equity and ecological validity. 

All AI-generated utterances were subjected to the same logging and timestamping protocols as human messages, enabling unified temporal and sequential analysis. Importantly, message scheduling was independent of persona design; supportive and contrarian agents followed identical timing parameters, ensuring that any differences in group discourse arose from behavioural stance rather than participation rate. Together, these interactional mechanisms allowed controlled yet lifelike AI participation, facilitating rigorous examination of how social persona, rather than timing artefacts, shaped the micro-dynamics of argumentative collaboration. A separate pilot with 15 participants (each paired with two AI teammates) produced 30 humanness ratings on a 7-point Likert scale, ranging from 1 (strongly disagree) to 7 (strongly agree). The results (M = 5.43, SD = 1.07) showed that AI teammates appeared convincingly human-like, supporting the ecological validity of the scheduling and interaction mechanisms. This pilot was conducted independently and was not included in the main sample or analysis.

\subsection{Experimental Design}

\begin{figure}[hbpt]
    \centering
    \includegraphics[width=0.75\linewidth]{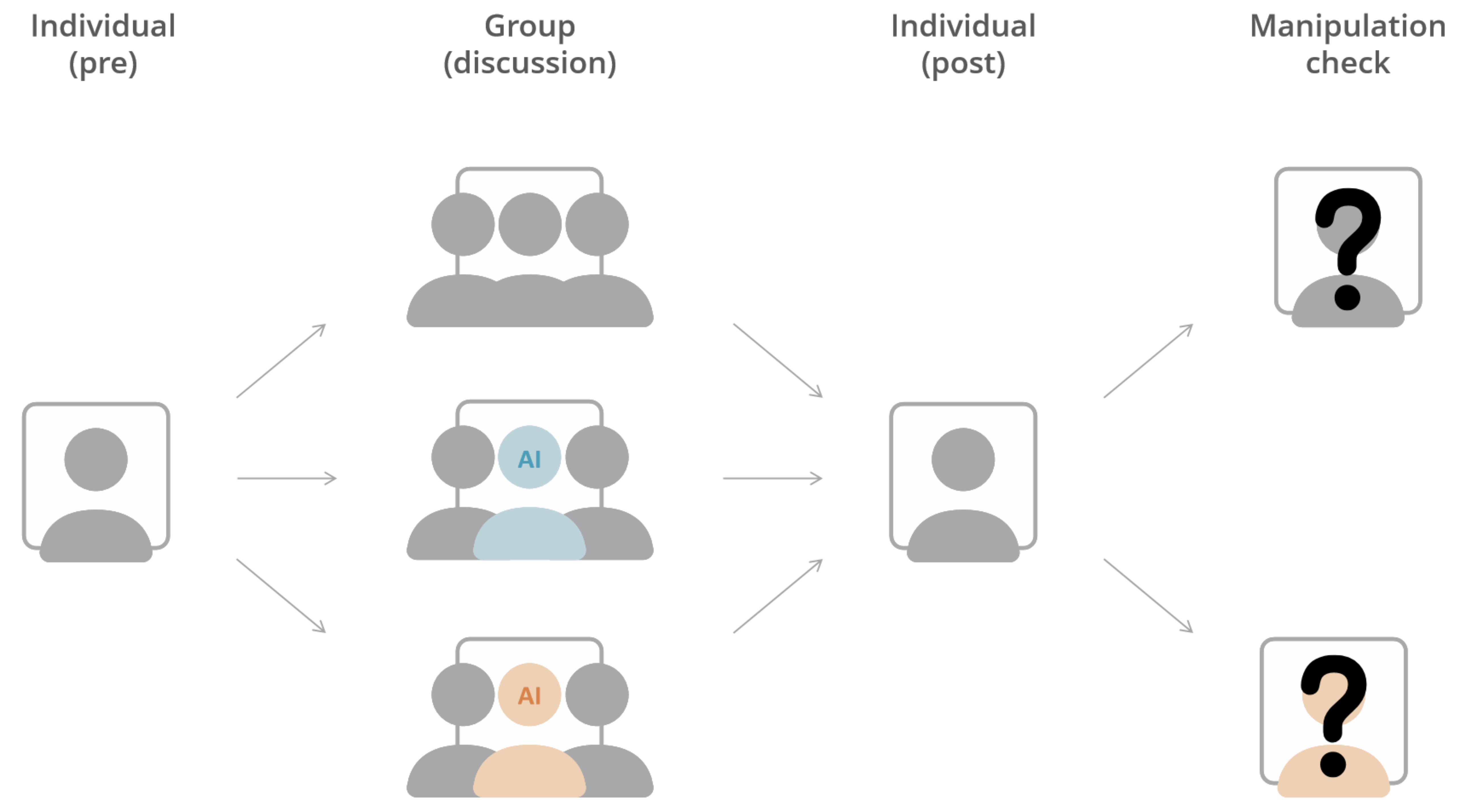}
    \caption{Overview of the experimental design implementing an \textit{individual-group-individual} (IGI) sequence. In Phase~A (\textit{Individual pre}), participants completed the survival-ranking task individually to establish a baseline. In Phase~B (\textit{Group discussion}), they collaborated in triads under one of three conditions: human-only control, supportive AI teammate, or contrarian AI teammate. In Phase~C (\textit{Individual post}), participants repeated the task individually to assess learning gains. A subsequent manipulation check assessed participants’ awareness of the AI teammate’s presence. Random assignment occurred at the group level, and participants were not informed of the AI’s involvement until post-task debriefing.}

    \label{fig:experimental-design}
\end{figure}

The study implemented a between-subjects experimental design to examine how the inclusion of a generative AI teammate influences collaborative reasoning during an analytical survival-ranking task (Figure \ref{fig:experimental-design}). Two main conditions were compared: (1) a human-only control condition, in which three human participants collaborated; and (2) a hybrid human-AI condition, in which two human participants collaborated with one generative AI agent. Within the hybrid condition, AI teammates were randomly assigned one of two personas, \textit{supportive} or \textit{contrarian}, representing contrasting social-epistemic orientations derived from collaborative learning theory. This design enabled comparison across three group types: Control, Supportive AI, and Contrarian AI. Collaboration followed an \textit{individual-group-individual} (\textit{IGI}) sequence. In \textit{Phase A}, participants completed the survival-ranking task individually, providing a baseline measure of reasoning and task performance. In \textit{Phase B}, triads engaged in a 10-minute synchronous group discussion to reach a consensus ranking; this phase generated the discourse corpus analysed in the present study. In \textit{Phase C}, participants repeated the task individually to measure post-discussion performance gains. Only the group-phase interactions were used for discourse and network analyses. Random assignment was conducted at the group level. Participants were informed that they would collaborate with two other online “students” in an online discussion task, without any explicit mention of whether their teammates were human or AI. No instructions were given to detect or distinguish between human and AI participants, as the study aimed to observe natural interaction dynamics within a typical online collaboration setting \citep{shanahan2023role}. Following task completion, all participants were fully debriefed regarding the inclusion and role of the AI collaborator.

\subsection{Measures}
\subsubsection{Discourse Coding and Reliability}
\label{sec:coding}

Group discussions were transcribed automatically from chat logs and segmented into individual utterances, each representing a single turn of speech. Every utterance was coded along the three dimensions from the \citet{Weinberger2006} framework: \textit{epistemic reasoning}, \textit{argument structure}, and \textit{social modes of co-construction}. The coding captured distinct but interrelated facets of argumentative knowledge construction, allowing analysis of both cognitive and interactional aspects of discourse (Table \ref{tab:coding_scheme}). Specifically, the \textit{epistemic dimension} captured the conceptual depth of reasoning through six mutually exclusive codes: Problem Space (EP-PS), Conceptual Space (EP-CS), Adequate Concept-Problem Relation (EP-CP-Adeq), Inadequate Concept-Problem Relation (EP-CP-Inad), Prior Knowledge (EP-PK), and Off-task (EP-OFF). The \textit{argument structure dimension} assessed the presence and quality of claims using five categories: Simple Claim (AR-C), Qualified Claim (AR-Cq), Grounded Claim (AR-Cg), Grounded and Qualified Claim (AR-Cqg), and Non-argumentative (AR-NA). Finally, the \textit{social modes of co-construction dimension} identified interactional functions of utterances as Externalisation (SOC-EXT), Elicitation (SOC-ELI), Quick Consensus (SOC-QCB), Integration-oriented (SOC-INT), or Conflict-oriented (SOC-CON).

\begin{table}[hbpt]
\centering
\caption{Coding scheme adopted from \citet{Weinberger2006} illustrating epistemic reasoning, argument structure, and social modes of co-construction}
\label{tab:coding_scheme}
\renewcommand{\arraystretch}{1.2}
\begin{tabular}{p{3cm}p{12.5cm}}
\toprule
\textbf{Code} & \textbf{Operational Definition} \\
\midrule
\multicolumn{2}{l}{\textbf{Epistemic reasoning}} \\[2pt]
EP-PS & Relates case facts to case facts; re-states scenario without theory. \\
EP-CS & Defines, contrasts, or elaborates theoretical concepts or heuristics independent of the case. \\
EP-CP-Adeq & Correctly applies a principle to the task situation, linking concept and case appropriately. \\
EP-CP-Inad & Misapplies a concept or demonstrates flawed reasoning. \\
EP-PK & Applies intuitive or everyday notions rather than domain-specific reasoning. \\
EP-OFF & Non-content or coordination utterances unrelated to task reasoning. \\
\midrule
\multicolumn{2}{l}{\textbf{Argument structure}} \\[2pt]
AR-C & States a position without qualifiers or supporting grounds. \\
AR-Cq & Provides a claim with a limitation or hedge. \\
AR-Cg & Presents a claim supported by explicit justification or evidence. \\
AR-Cqg & Combines both warrant and qualifier within the same statement. \\
AR-NA & Non-claim statements such as coordination or questions. \\
\midrule
\multicolumn{2}{l}{\textbf{Social modes of co-construction}} \\[2pt]
SOC-EXT & Introduces a new idea without reference to others’ contributions. \\
SOC-ELI & Requests information, feedback, or action from peers. \\
SOC-QCB & Expresses agreement or paraphrase without modification. \\
SOC-INT & Extends or refines a peer’s reasoning by adding new conditions or merging criteria. \\
SOC-CON & Challenges or replaces a peer’s point through disagreement or alternative proposal. \\
\bottomrule
\end{tabular}
\end{table}

Each utterance received exactly one code per dimension. Coding was conducted by two trained human raters with expertise in collaborative learning and discourse analysis. Prior to formal coding, both raters completed a calibration phase involving joint annotation of 20\% of the utterances (632 utterances) to establish a shared understanding of the code definitions and decision rules. Coding guidelines were iteratively refined through discussion until full conceptual alignment was reached. During the main coding phase, each rater independently coded the same subset of transcripts, and regular cross-checking sessions were held to resolve discrepancies and maintain consistency. Coders examined each message within a contextual window of three preceding and three subsequent utterances to ensure semantic coherence when assigning codes. Additionally, coders were blind to the group condition to prevent bias. The final dataset comprised 3{,}160 coded utterances across all groups. 

Inter-rater reliability was assessed by two independent coders on the full sample of the 3{,}160 utterances. Cohen’s~$\kappa$ coefficients indicated consistently high levels of agreement across all dimensions, with values ranging from substantial to almost perfect. Reliability was strongest for the Social Modes of Co-construction ($\kappa = 0.920$), followed by Argument Structure ($\kappa = 0.900$) and Epistemic Reasoning ($\kappa = 0.866$). The overall reliability across all categories was $\kappa = 0.784$, exceeding the conventional threshold of~0.70 for acceptable agreement \citep{Landis1977}. Table~\ref{tab:reliability} presents detailed reliability statistics for every code within each dimension, including corresponding frequencies. All subcodes achieved at least substantial reliability. Complete confusion matrices and cross-coder comparison tables are available in the Appendix \ref{appendix-confusion}.

\begin{table}[ht]
\centering
\caption{Inter-rater reliability of discourse coding across dimensions and codes ($N=3160$ utterances)}
\label{tab:reliability}
\begin{tabular}{llccc}
\toprule
\textbf{Dimension} & \textbf{Code} & \textbf{Cohen’s $\boldsymbol{\kappa}$} & \textbf{Interpretation} & \textbf{N (\%)} \\
\midrule
\multirow{7}{*}{\textbf{Epistemic reasoning}} 
 & EP-PS & 0.852 & Almost perfect & 1197 (37.9) \\
 & EP-CS & 0.823 & Almost perfect & 17 (0.5) \\
 & EP-CP-Adeq & 0.877 & Almost perfect & 440 (13.9) \\
 & EP-CP-Inad & 0.759 & Substantial & 171 (5.4) \\
 & EP-PK & 0.812 & Almost perfect & 574 (18.2) \\
 & EP-OFF & 0.954 & Almost perfect & 761 (24.1) \\
 & \textit{Overall (Epistemic)} & \textbf{0.866} & \textbf{Almost perfect} & \textbf{3160 (100)} \\
\midrule
\multirow{6}{*}{\textbf{Argument structure}} 
 & AR-C & 0.856 & Almost perfect & 401 (12.7) \\
 & AR-Cq & 0.831 & Almost perfect & 333 (10.5) \\
 & AR-Cg & 0.929 & Almost perfect & 836 (26.5) \\
 & AR-Cqg & 0.724 & Substantial & 148 (4.7) \\
 & AR-NA & 0.955 & Almost perfect & 1441 (45.6) \\
 & \textit{Overall (Argument)} & \textbf{0.900} & \textbf{Almost perfect} & \textbf{3160 (100)} \\
\midrule
\multirow{6}{*}{\textbf{Social modes of co-construction}} 
 & SOC-EXT & 0.925 & Almost perfect & 1039 (32.9) \\
 & SOC-ELI & 0.958 & Almost perfect & 562 (17.8) \\
 & SOC-QCB & 0.925 & Almost perfect & 611 (19.3) \\
 & SOC-INT & 0.842 & Almost perfect & 366 (11.6) \\
 & SOC-CON & 0.923 & Almost perfect & 582 (18.4) \\
 & \textit{Overall (Social modes)} & \textbf{0.920} & \textbf{Almost perfect} & \textbf{3160 (100)} \\
\midrule
\textbf{All dimensions combined} & \textit{Overall reliability} & \textbf{0.784} & \textbf{Substantial} & \textbf{3160 (100)} \\
\bottomrule
\end{tabular}
\end{table}

\subsubsection{Quantitative Indices}

To address RQ1 and RQ5, we computed four quantitative indices based on the \citet{Weinberger2006} framework, corresponding to the core process dimensions of argumentative knowledge construction: participation, epistemic adequacy, argument quality, and transactivity. These indices served as predictors in the subsequent linear mixed-effects models (LMMs). All indices were calculated at the \textit{individual level} for human participants and then aggregated to the group level for descriptive comparison; individual-level values were retained in modelling to preserve within-group variance. Each continuous index was $z$-standardised prior to analysis. Proportion-based indices (EA and AQI) were additionally logit-transformed in robustness checks to normalise bounded distributions, whereas PI and TI were modelled on the standardised continuous scale.

The \textit{Participation} dimension captures two complementary aspects of learners’ engagement in collaborative discourse: the quantity of participation and the heterogeneity of participation. The quantity of participation ($\text{Quantity}_{ig}$) represents the extent of each learner’s verbal contribution to the discussion. For each participant $i$ in group $g$ with $J_g$ members, let $m_{ig}$ denote the total number of words contributed during the collaborative task. Because word counts are typically right-skewed, they were log-transformed to reduce heteroscedasticity and improve model interpretability. This raw, log-transformed participation measure preserves between-group variance and enables estimation of condition effects in the LMM analysis.

\begin{equation}
\label{eq:PI_quantity}
\text{Quantity}_{ig} = \log (m_{ig})
\end{equation}

The heterogeneity of participation ($\text{Equality}_{g}$) reflects how evenly discourse contributions were distributed among group members. To assess this equality, the Gini coefficient of word counts was calculated for each group, with higher Gini values indicating greater inequality. The equality score was then expressed as the complement of the Gini coefficient, such that higher values represent more balanced participation. This measure ranges from~0 (perfect inequality) to~1 (perfect equality) and provides a group-level indicator of equitable engagement.

\begin{equation}
\label{eq:PI_equality}
\text{Equality}_{g} = 1 - \mathrm{Gini}\!\left(\{m_{1g}, m_{2g}, \dots, m_{J_g g}\}\right)
\end{equation}

The \textit{Epistemic Adequacy Ratio (EA)} operationalises the epistemic dimension, reflecting the proportion of conceptually adequate reasoning relative to all epistemic activities (Eq.~\ref{eq:EA}). Adequate concept-problem relations (\text{EP-CP-Adeq}) represent the most integrative form of reasoning, where conceptual understanding is correctly applied to the problem context. Higher $\text{EA}_{ig}$ values thus indicate more conceptually grounded and contextually appropriate reasoning.

\begin{equation}
\label{eq:EA}
\text{EA}_{ig} =
\frac{\text{EP-CP-Adeq}_{ig}}
{\text{EP-CP-Adeq}_{ig} + \text{EP-CP-Inad}_{ig} + \text{EP-PK}_{ig} + \text{EP-PS}_{ig} + \text{EP-CS}_{ig}}
\end{equation}

The \textit{Argument Quality Index (AQI)} captures the epistemic sophistication of argumentation (Eq.~\ref{eq:AQI}). It expresses the proportion of higher-order claims, those that are both grounded in justification (\text{AR-Cg}) and/or qualified (\text{AR-Cqg}), relative to all claims produced. This ratio reflects the degree to which learners supported and qualified their assertions, with higher $\text{AQI}_{ig}$ values signifying more elaborated and evidence-based reasoning.

\begin{equation}
\label{eq:AQI}
\text{AQI}_{ig} =
\frac{\text{AR-Cg}_{ig} + \text{AR-Cqg}_{ig}}
{\text{AR-C}_{ig} + \text{AR-Cq}_{ig} + \text{AR-Cg}_{ig} + \text{AR-Cqg}_{ig}}
\end{equation}

The \textit{Transactivity Index (TI)} reflects the social co-construction dimension, representing how actively participants build on, integrate, or critique others’ reasoning (Eq.~\ref{eq:TI}). For each individual $i$ in group $g$, the frequencies of the five social modes, externalisation (\text{SOC-EXT}), elicitation (\text{SOC-ELI}), quick-consensus (\text{SOC-QCB}), integration-oriented (\text{SOC-INT}), and conflict-oriented (\text{SOC-CON}), were weighted ordinally from 1 to 5 to represent increasing transactivity \citep{Weinberger2006}. The weighted mean yields a continuous measure where higher $\text{TI}_{ig}$ values indicate more transactive, cognitively engaged collaboration.

\begin{equation}
\label{eq:TI}
\text{TI}_{ig} =
\frac{1\!\cdot\!\text{SOC-EXT}_{ig} + 2\!\cdot\!\text{SOC-ELI}_{ig} + 3\!\cdot\!\text{SOC-QCB}_{ig} + 4\!\cdot\!\text{SOC-INT}_{ig} + 5\!\cdot\!\text{SOC-CON}_{ig}}
{\text{SOC-EXT}_{ig} + \text{SOC-ELI}_{ig} + \text{SOC-QCB}_{ig} + \text{SOC-INT}_{ig} + \text{SOC-CON}_{ig}}
\end{equation}

For the outcome measure, we compute the \textit{Task Performance Score (PS)}, which provides an objective measure of individual problem-solving accuracy in the \textit{Winter Survival} task at each time point (Eq.~\ref{eq:PS}). Each participant produced an individual ranking of 12 survival items before group discussion ($\text{PS}_{\text{pre},i}$) and again after discussion ($\text{PS}_{\text{post},i}$). For both assessments, the score was calculated as the sum of absolute deviations between the participant’s ranking and the expert reference ranking (see Apendix \ref{appendix-task}), with smaller values indicating greater alignment with expert judgment and thus higher task performance. This scoring method yields a continuous index of individual reasoning accuracy, allowing comparison between pre- and post-discussion performance as indicators of learning or convergence toward expert consensus.

\begin{equation}
\label{eq:PS}
\text{PS}_{t,i} = \sum_{k=1}^{12} \left| R_{i,k,t} - R^{*}_{k} \right|
\end{equation}

\subsection{Analysis}

\subsubsection{Preliminary: Manipulation Checks}

To ensure the validity of the AI manipulation and control for potential confounds, three post-hoc checks were conducted. \textit{AI sensitivity} was measured through a post-task questionnaire in which participants identified whether each teammate (\texttt{Kevin}, \texttt{Stuart}, \texttt{Bob}) was human, AI, or if they were unsure, enabling assessment of awareness and potential misattribution of AI identity. Detection accuracy was computed as the proportion of correct identifications relative to the total number of teammate judgments, with “Not sure” responses treated as inaccurate, adhering to the best practice in detection theory \citep{lynn2014utilizing}. \textit{AI talkativeness} was examined using system logs that recorded all chat messages, from which we calculated the average number of messages and mean words per message for each persona to confirm comparable verbosity across supportive and contrarian AIs. \textit{Persona fidelity} was assessed through both linguistic and perceptual measures. Linguistically, a LIWC-22 analysis of AI utterances focused on the “Emotional Tone” dimension as an indicator of stance, where higher scores reflected more positive and affiliative language (supportive persona) and lower scores indicated a more critical or challenging style (contrarian persona). Perceptually, participants responded to two post-survey items evaluating the perceived social stance of each teammate on a 7-point Likert scale (1 = \textit{strongly disagree}, 7 = \textit{strongly agree}): “Kevin/Stuart/Bob was supportive during the group discussion” and “Kevin/Stuart/Bob was contrarian during the group discussion.” These perceptual ratings provided behavioural validation of the intended persona manipulation. For all three manipulation checks, descriptive statistics and Mann-Whitney~$U$~tests were reported to compare the supportive and contrarian conditions, providing non-parametric verification of equivalence or intended differences where appropriate.

\subsubsection{RQ1: Participation Analysis}

To address RQ1, the effects of AI teammates on learners’ participation were analysed using two complementary indices: the \textit{quantity of participation} ($\text{Quantity}_{ig}$) and the \textit{heterogeneity of participation} ($\text{Equality}_{g}$). A linear mixed-effects model (LMM; Eq.~\ref{eq:LMM_RQ1}) was used to examine differences in $\log$-transformed participation quantity across the three experimental conditions (Control, Supportive AI, Contrarian AI), with \textit{Condition} entered as a fixed effect and \textit{Group ID} included as a random intercept to account for the nested structure of individuals within groups. Because $\text{Equality}_{g}$ is a bounded (0-1) index derived from the complement of the Gini coefficient and may violate normality assumptions, differences across conditions were evaluated using a non-parametric approach. Specifically, a Kruskal-Wallis test was conducted to examine overall condition effects, followed by Dunn’s pairwise comparisons with Holm $p$-value adjustment to control the familywise error rate. This distribution-free approach provides a robust group-level assessment of participation equality across experimental conditions.

\begin{equation}
\label{eq:LMM_RQ1}
\log(\text{Quantity}_{ig}) = \beta_0 + \beta_1\text{Condition}_{g} + u_{g} + \epsilon_{ig}
\end{equation}

\subsubsection{RQ2--4: Epistemic Network Analysis}

To address RQ2--4, epistemic network analysis (ENA; \citealp{shaffer2016tutorial}) was conducted to model the structure and strength of connections among discourse codes within each dimension of the Weinberger-Fischer framework. ENA provides a quantitative representation of how categories co-occur in temporal proximity, enabling visual and statistical comparison of discourse patterns across experimental conditions. Separate ENA models were constructed for each discourse dimension, epistemic reasoning, argument structure, and social modes of co-construction. The unit of analysis was defined as the combination of experimental condition and group identifier (\textit{condition × group\_id}), treating each group discussion as an independent case. Co-occurrences were computed within a sliding window of four utterances (looking four turns backward), capturing short-range temporal dependencies between codes (following the recommended guideline \citep{shaffer2016tutorial}). Mean rotation was applied to maximise variance along the first principal axis (MR1) representing the dominant contrast between conditions, while orthogonally projecting residual variance onto a second dimension (SVD2; \citealp{shaffer2016tutorial}). Each group’s position in the two-dimensional ENA space reflected the relative emphasis of co-occurring discourse elements within that group’s interaction.

Subtraction networks were generated to visualise differences between conditions. Three pairwise comparisons were examined: (1) Supportive AI versus Control, (2) Contrarian AI versus Control, and (3) Contrarian AI versus Supportive AI. In these difference networks, blue edges indicated stronger co-occurrence in the first condition, and red edges indicated stronger co-occurrence in the second. Edge thickness corresponded to the magnitude of the difference in mean edge weights between groups. For clarity, edge weights were scaled uniformly (1.5×) and colours standardised across all plots. Group centroid positions on the MR1 and SVD2 axes were compared using Mann-Whitney $U$ tests (non-parametric) for each pairwise contrast. Effect sizes were reported using Cliff’s~$\delta$, which is robust to non-normal distributions, and interpreted following \citet{Romano2006}: negligible ($|\delta|<0.147$), small ($0.147\leq|\delta|<0.330$), medium ($0.330\leq|\delta|<0.474$), and large ($|\delta|\geq0.474$). Statistical significance was assessed at $\alpha = .05$, with Bonferroni-adjusted thresholds applied within each dimension ($\alpha/3 = .0167$).

\subsubsection{RQ5: Linear Mixed-Effects Models}

To address RQ5, we used \textit{linear mixed-effects models} (LMMs) to examine how discourse processes predicted post-discussion task performance at the individual level. Given the nested structure of participants ($i$) within collaborative groups ($g$), LMMs appropriately account for the non-independence of individuals working in the same team and enable estimation of both within- and between-group effects. The dependent variable was the \textit{post-discussion task performance score} ($\text{PS}_{\text{post},ig}$), representing each participant’s accuracy in ranking survival items after collaboration, with the corresponding pre-discussion score ($\text{PS}_{\text{pre},ig}$) included as a covariate to control for baseline differences. Lower PS values indicate higher performance accuracy, as the score reflects the total deviation from the expert benchmark (Appendix~\ref{appendix-task}). Due to technical errors in recording post-task responses, data from nine participants were excluded, resulting in 203 valid participants nested within 92 groups: 83 in the human-only condition (28 groups), 60 in the supportive AI condition (32 groups), and 60 in the contrarian AI condition (32 groups). All model residuals and diagnostic plots confirmed normality, homoscedasticity, and absence of influential outliers, and the intercorrelation matrix among predictors showed no multicollinearity concerns; detailed diagnostics and correlation matrices are provided in Appendix~\ref{appendix-diagnostics}.

\begin{equation}
\label{eq:LMM-DV}
\text{PS}_{\text{post},ig} = f(\text{PS}_{\text{pre},ig}, \text{Quantity}_{ig}, \text{EA}_{ig}, \text{AQI}_{ig}, \text{TI}_{ig}, \text{Condition}_{g}, \text{Equality}_{g})
\end{equation}

At Level~1 (individual level), the model captured how each participant’s discourse processes predicted post-discussion performance, controlling for their baseline score (Eq.~\ref{eq:LMM-L1}). Specifically, individual-level predictors included the standardised quantity of participation $z(\text{Quantity}_{ig})$, epistemic adequacy (EA), argument quality (AQI), and transactivity (TI), in addition to the pre-discussion task performance $\text{PS}_{\text{pre},ig}$. Here, $\beta_{0g}$ represents the group-specific intercept, $\beta_{k}$ the fixed slopes for each discourse predictor, and $r_{ig}$ the individual-level residual, assumed to be normally distributed $r_{ig} \sim N(0, \sigma^2)$.

\begin{equation}
\label{eq:LMM-L1}
\text{PS}_{\text{post},ig} = \beta_{0g} + \beta_{1}\text{PS}_{\text{pre},ig} + \beta_{2}z(\text{Quantity}_{ig}) + \beta_{3}\text{EA}_{ig} + \beta_{4}\text{AQI}_{ig} + \beta_{5}\text{TI}_{ig} + r_{ig}
\end{equation}

At Level~2 (group level), variation in intercepts and slopes was modelled as a function of collaborative condition (i.e., AI teammate type) and the equality of participation ($\text{Equality}_{g}$). In this specification, $\gamma_{00}$ denotes the grand mean intercept, $\gamma_{01}$ and $\gamma_{02}$ the fixed effects of condition and equality, respectively, and $u_{0g}$ and $u_{kg}$ the group-level random effects, each assumed to follow $N(0, \tau^2)$ (Eq.~\ref{eq:LMM-L2}).

\begin{equation}
\label{eq:LMM-L2}
\begin{aligned}
\beta_{0g} &= \gamma_{00} + \gamma_{01}(\text{Condition}_{g}) + \gamma_{02}(\text{Equality}_{g}) + u_{0g},\\
\beta_{k}  &= \gamma_{k0} + u_{kg}, \quad k = 1,\dots,5
\end{aligned}
\end{equation}

The combined model (Eq.~\ref{eq:LMM-combined}) therefore estimated how individual-level discourse indicators and group-level factors jointly predicted post-discussion task performance, while controlling for baseline ability and accounting for shared variance within groups. 

\begin{equation}
\label{eq:LMM-combined}
\text{PS}_{\text{post},ig} = \gamma_{00} + \gamma_{01}(\text{Condition}_{g}) + \gamma_{02}(\text{Equality}_{g}) 
+ \gamma_{10}\text{PS}_{\text{pre},ig} + \gamma_{20}z(\text{Quantity}_{ig}) + \gamma_{30}\text{EA}_{ig} 
+ \gamma_{40}\text{AQI}_{ig} + \gamma_{50}\text{TI}_{ig} 
+ u_{0g} + r_{ig}
\end{equation}

All models were estimated using restricted maximum likelihood (REML) with cluster-robust (CR2) standard errors to correct for small-sample bias and unbalanced group sizes. A sequential model comparison procedure was conducted: (a) a null model to compute the intraclass correlation coefficient (ICC), (b) a random-intercept model with individual-level predictors (Model 1), (c) a full model including group-level predictors and cross-level interactions (Model 2), and (d) a robustness model using logit-transformed, $z$-standardised proportions for bounded predictors (EA and AQI) to verify coefficient stability (Table \ref{tab:lmm_config}). Model fit was evaluated using the $-2$ log-likelihood, Akaike Information Criterion (AIC), and Bayesian Information Criterion (BIC). These analyses collectively assessed how participation quantity, participation equality, epistemic reasoning, argumentation, and transactivity predicted post-discussion performance while accounting for between-group variation.

\begin{table}[ht]
\centering
\caption{Configurations of linear mixed-effects models for RQ5}
\label{tab:lmm_config}
\renewcommand{\arraystretch}{1.2}
\begin{tabular}{p{2.5cm}p{12.5cm}}
\toprule
\textbf{Model} & \textbf{Specification and Included Predictors} \\
\midrule
\textbf{Null Model} & Random-intercept model including pre-task performance as a control variable to estimate the intraclass correlation coefficient (ICC). Captures the proportion of variance in post-task performance attributable to group membership. \\
\addlinespace
\textbf{Model 1} & 
Includes baseline performance and individual-level discourse indices: participation quantity, epistemic adequacy, argument quality, and transactivity: \newline
$\text{PS}_{\text{post},ig} \sim \text{PS}_{\text{pre},ig} + z(\text{Quantity}_{ig}) + \text{EA}_{ig} + \text{AQI}_{ig} + \text{TI}_{ig} + (1|\text{Group})$  \\
\addlinespace
\textbf{Model 2} & 
Extends Model 1 by adding collaborative condition (Control, Supportive AI, Contrarian AI) and group-level participation equality to account for between-group variation: \newline
$\text{PS}_{\text{post},ig} \sim \text{PS}_{\text{pre},ig} + z(\text{Quantity}_{ig}) + \text{EA}_{ig} + \text{AQI}_{ig} + \text{TI}_{ig} + \text{Condition}_{g} + z(\text{Equality}_{g}) + (1|\text{Group})$ \\
\addlinespace
\textbf{Robustness check} & 
Re-estimation of Model 1 using logit-transformed, $z$-standardised proportions for EA and AQI to normalise bounded distributions and verify coefficient stability. \\
\bottomrule
\end{tabular}
\end{table}

\section{Results}

\subsection{Manipulation Checks}

Manipulation checks confirmed that the AI personas were successfully differentiated while maintaining comparable levels of overall participation. Participants demonstrated similar \textit{AI sensitivity} across conditions, with correct identification rates of 28.8\%~[18.0, 41.9] for the Contrarian AI and 28.3\%~[17.4, 41.4] for the Supportive AI, indicating limited awareness of the agents’ artificial identity. Linguistic analyses revealed that the Supportive AI (Median~=~90.42, IQR~=~13.15) exhibited a substantially more positive emotional tone than the Contrarian AI (Median~=~38.53, IQR~=~35.16), $U~=~$, $p~<~.001$, $\Delta_{\text{Cliff}}~=~-0.975$, reflecting a \textit{very large} effect size. 

Behavioural analyses further indicated that both personas were comparably active in message frequency and total word output. The number of messages did not differ significantly between conditions (Supportive: Median~=~10.0, IQR~=~9.0--13.0; Contrarian: Median~=~11.0, IQR~=~10.8--13.0), $U~=~397.5$, $p~=~.121$, $\Delta_{\text{Cliff}}~=~-0.224$ (small). Similarly, total words produced were equivalent across personas (Supportive: Median~=~180.5, IQR~=~169.0--208.2; Contrarian: Median~=~183.5, IQR~=~160.0--195.5), $U~=~531.5$, $p~=~.799$, $\Delta_{\text{Cliff}}~=~0.038$ (negligible). However, a significant difference emerged in the average number of words per message: the Supportive AI (Median~=~17.72, IQR~=~16.00--18.56) wrote longer messages than the Contrarian AI (Median~=~15.58, IQR~=~14.48--17.28), $U~=~743.0$, $p~=~.002$, $\Delta_{\text{Cliff}}~=~0.451$ (medium), consistent with its elaborative and affiliative communication style. 

Perceptual ratings corroborated these distinctions: participants rated the Supportive AI as more \textit{supportive} (Median~=~6.0, IQR~=~1.5) than the Contrarian AI (Median~=~5.0, IQR~=~2.0), $U~=~1055.5$, $p~<~.001$, $\Delta_{\text{Cliff}}~=~-0.394$ (medium), and rated the Contrarian AI as more \textit{contrary} (Median~=~4.0, IQR~=~2.5) than the Supportive AI (Median~=~2.0, IQR~=~2.0), $U~=~2509.5$, $p~<~.001$, $\Delta_{\text{Cliff}}~=~0.442$ (medium). Collectively, these results demonstrate that the two AI personas differed systematically in tone, stance, and linguistic elaboration, while maintaining equivalent levels of verbosity and message frequency, confirming a successful manipulation of social-epistemic orientation without confounding effects of participation quantity.

\subsection{RQ1: Participation}

Analyses examined whether AI teammates influenced learners’ participation quantity and equality during collaborative problem solving. At the individual level, non-parametric comparisons of human participants’ discourse activity revealed no significant differences across experimental conditions. The number of messages was similar among groups (Control: Median~=~10.0, IQR~=~8.0--12.0; Supportive AI: Median~=~12.0, IQR~=~8.0--14.0; Contrarian AI: Median~=~11.0, IQR~=~8.0--15.0), with no significant effect of condition, $H~=~2.67$, $p~=~.263$. Likewise, the average words per message did not differ significantly across conditions (Control: Median~=~10.77, IQR~=~7.52--14.12; Supportive AI: Median~=~9.94, IQR~=~8.23--11.98; Contrarian AI: Median~=~9.19, IQR~=~8.00--11.77), $H~=~3.20$, $p~=~.202$, with only a small, non-significant trend between the Control and Contrarian groups ($\Delta_{\text{Cliff}}~=~0.170$, $p_{\text{adj}}~=~.230$). Likewise, at the group level, participation equality, measured as the complement of the Gini coefficient, also showed no significant condition effects ($H~=~0.76$, $p~=~.684$). Median equality scores were highly similar across conditions (Control: Median~=~0.871, IQR~=~0.819--0.938; Supportive AI: Median~=~0.890, IQR~=~0.800--0.936; Contrarian AI: Median~=~0.881, IQR~=~0.840--0.946), indicating balanced participation within all groups. Overall, the presence and persona of AI teammates did not significantly affect either the quantity or the equality of learners’ contributions, suggesting that engagement levels remained stable across all experimental conditions.

\subsection{RQ2: Epistemic Reasoning}

For the epistemic reasoning dimension, the first and second ENA dimensions explained 23.68\% and 29.77\% of the variance, respectively, capturing the dominant contrasts in epistemic co-occurrence patterns across groups. Mann-Whitney tests revealed significant, large effects on the first dimension (MR1) when comparing both AI conditions to the human-only Control group. Supportive AI groups scored lower on MR1 than Control ($U~=~161$, $p~<~.001$, $\Delta_{\text{Cliff}}~=~-0.641$, large), whereas Contrarian AI groups scored higher ($U~=~736$, $p~<~.001$, $\Delta_{\text{Cliff}}~=~0.643$, large). No significant differences were observed between the two AI personas or along the second dimension (SVD2; all $p > .05$). These findings indicate that the presence of AI teammates, regardless of persona, substantially altered the configuration of epistemic connections compared to human-only groups.

\begin{figure}
    \centering
    \includegraphics[width=0.75\linewidth]{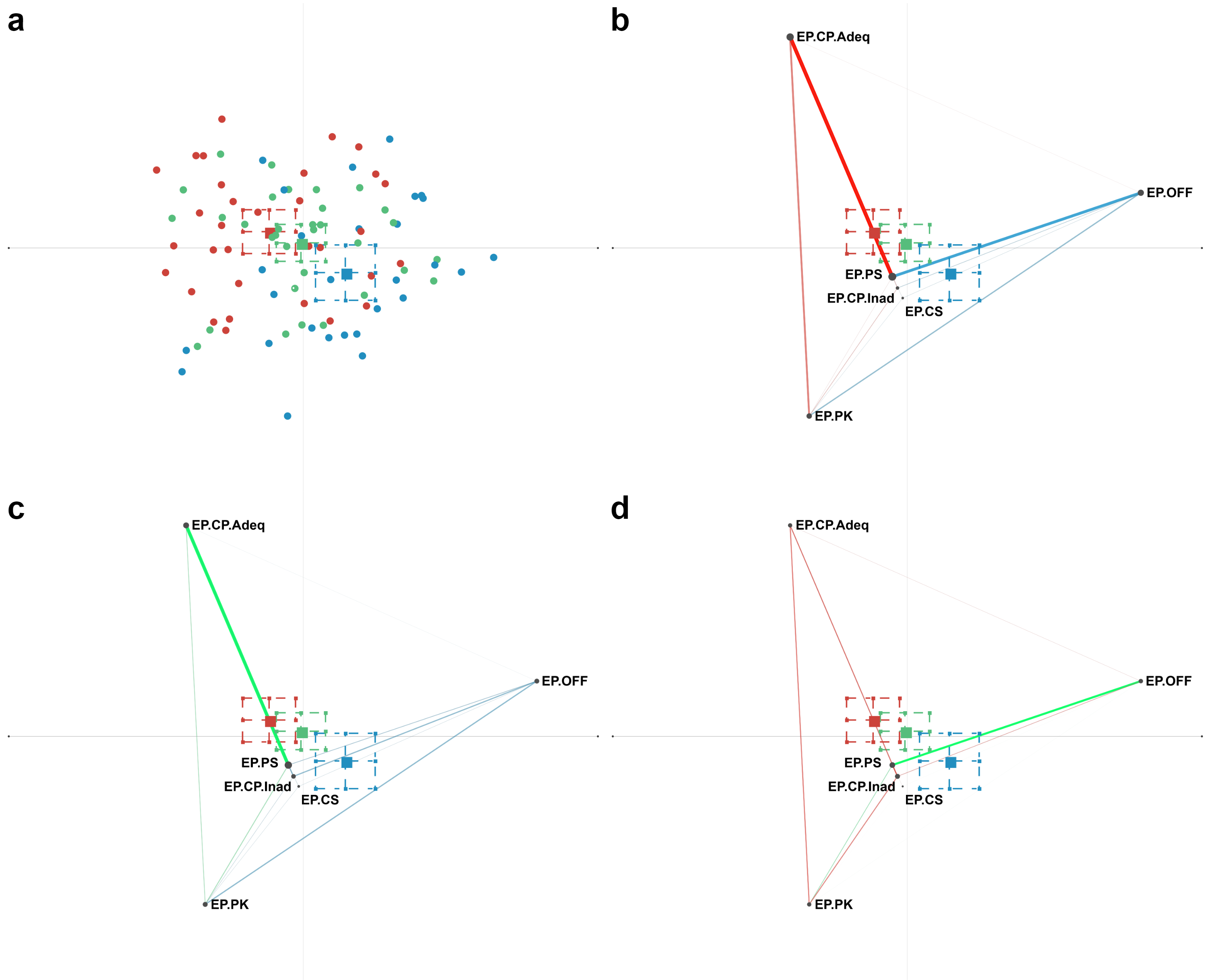}
    \caption{Epistemic Network Analysis (ENA) results for the \textit{epistemic reasoning} dimension. 
    (a) Group centroid distributions with 95\% confidence ellipses by condition. 
    (b) Contrarian AI (red) vs.~Control (blue), showing stronger epistemic links in each. 
    (c) Supportive AI (green) vs.~Control, highlighting greater conceptual integration. 
    (d) Contrarian vs.~Supportive AI, revealing contrasting problem- and concept-oriented reasoning patterns.}
    \label{fig:ena-ep}
\end{figure}

In the Supportive AI condition (Figure \ref{fig:ena-ep}), the most pronounced differences relative to Control were stronger co-occurrences between \textit{Problem Space} and \textit{Adequate Conceptual Progress} (EP-PS~\&~EP-CP-Adeq; weight~=~0.206), indicating that supportive agents helped consolidate reasoning around valid conceptual-problem linkages. By contrast, the Control condition showed comparatively weaker integration among these task-relevant categories. Similarly, Contrarian AI groups demonstrated stronger associations than Control between \textit{Problem Space} and \textit{Adequate Conceptual Progress} (EP-PS~\&~EP-CP-Adeq; weight~=~0.234) and between \textit{Adequate Conceptual Progress} and \textit{Prior Knowledge} (EP-CP-Adeq~\&~EP-PK; weight~=~0.116). These high-weight edges reflect that contrarian prompting fostered more frequent coupling of conceptual adequacy with prior knowledge and problem framing. Overall, AI teammates, particularly in supportive and contrarian roles, enhanced the structural integration of epistemic reasoning compared to human-only collaboration, though their effects differed in orientation rather than magnitude.

\subsection{RQ3: Argument Structure}

For the argument structure dimension, the ENA model accounted for 27.84\% of variance on the first dimension (MR1) and 32.38\% on the second (SVD2). Mann-Whitney tests revealed that Supportive AI groups differed significantly from Control on MR1 ($U~=~641$, $p~=~.004$, $\Delta_{\text{Cliff}}~=~0.431$, medium), whereas Contrarian AI groups did not ($p~=~.529$). However, Contrarian groups differed significantly from both Control and Supportive groups on SVD2 ($U~=~224$, $p~=~.001$, $\Delta_{\text{Cliff}}~=~-0.500$, large; $U~=~219$, $p~=~.001$, $\Delta_{\text{Cliff}}~=~-0.572$, large) and from Supportive groups on MR1 ($U~=~844$, $p~=~.001$, $\Delta_{\text{Cliff}}~=~0.648$, large). These findings indicate that the two AI personas shaped argumentation patterns along distinct structural dimensions: the Supportive AI promoted consistent claim elaboration, while the Contrarian AI induced deeper but more divergent argumentative linkages.

\begin{figure}
    \centering
    \includegraphics[width=0.75\linewidth]{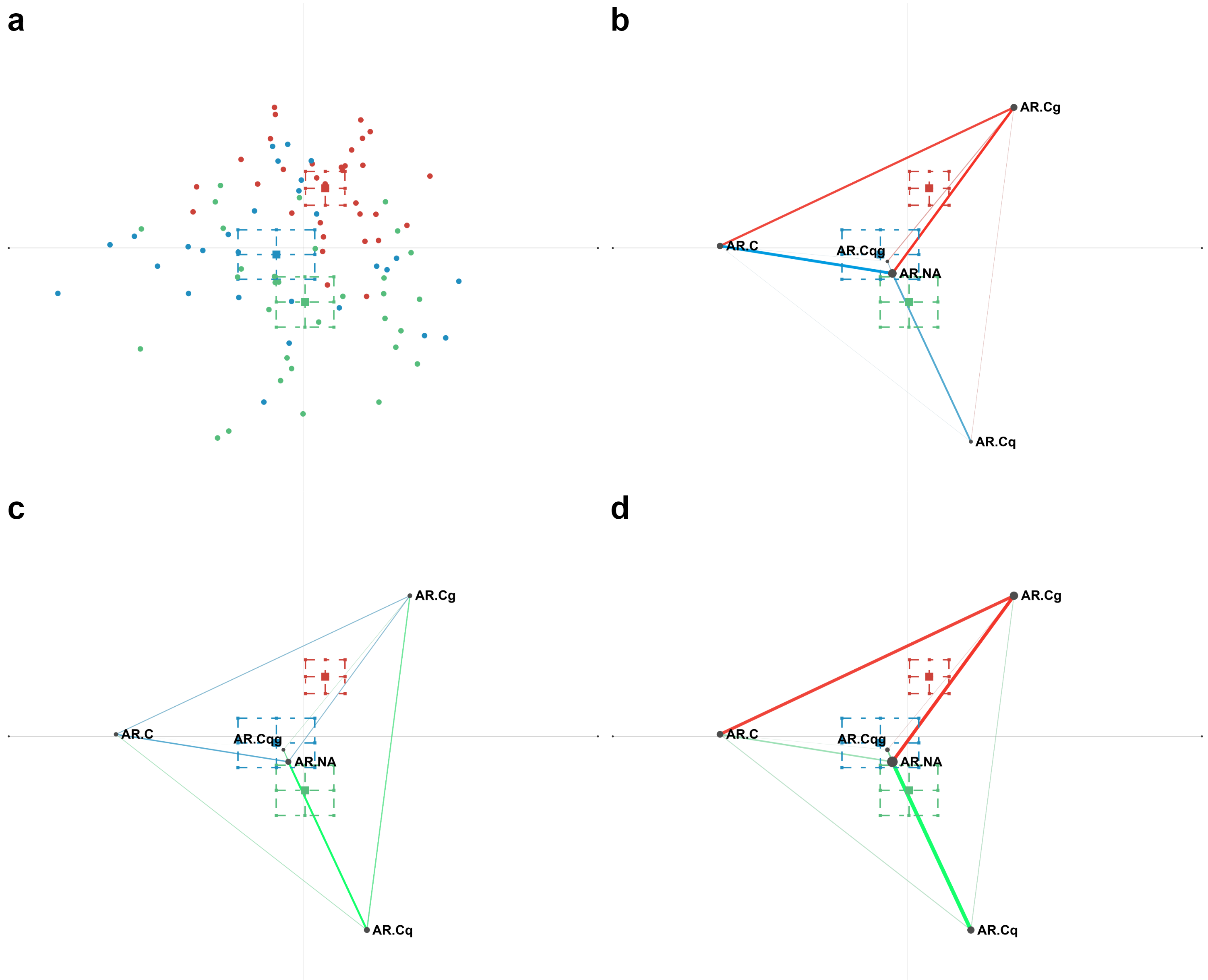}
    \caption{Epistemic Network Analysis (ENA) results for the \textit{argument structure} dimension. 
    (a) Group distributions in ENA space with mean centroids and 95\% confidence ellipses. 
    (b) Contrarian AI (red) vs.~Control (blue) subtraction network, highlighting stronger links among counterclaims and qualifiers in Contrarian groups. 
    (c) Supportive AI (green) vs.~Control, showing more frequent co-occurrences between claims and grounds in Supportive groups. 
    (d) Contrarian vs.~Supportive AI, revealing distinct argumentation styles driven by persona orientation.}
    \label{fig:ena-ar}
\end{figure}

At the network level (Figure \ref{fig:ena-ar}), Supportive AI groups displayed one high-weight contrast relative to Control, namely a stronger co-occurrence between \textit{qualified claims} (AR-Cq) and \textit{non-argumentative utterances} (AR-NA; weight~=~0.124), suggesting smoother transitions between reasoning and conversational scaffolding. Contrarian AI groups, by contrast, exhibited stronger connections than Control between \textit{claims with grounds} (AR-Cg) and \textit{non-argumentative utterances} (AR-NA; weight~=~0.153) and between \textit{claims} (AR-C) and \textit{claims with grounds} (AR-Cg; weight~=~0.137). These patterns indicate that Contrarian AI teammates encouraged participants to interleave direct claims with grounded elaborations, reflecting a more argumentative orientation. 

The comparison between Contrarian and Supportive AI conditions further revealed high-weight differences exceeding the reporting threshold. Contrarian groups showed denser linkages between \textit{claims with grounds} (AR-Cg) and \textit{non-argumentative statements} (AR-NA; weight~=~0.206), and between \textit{claims} (AR-C) and \textit{claims with grounds} (AR-Cg; weight~=~0.192), whereas Supportive groups displayed stronger connections between \textit{qualified claims} (AR-Cq) and \textit{non-argumentative utterances} (AR-NA; weight~=~0.237). Together, these results suggest that supportive agents facilitated fluid exchanges between everyday remarks and qualified statements, whereas contrarian agents fostered concentrated linkages between core claim structures and their evidential grounding, reflecting distinct, persona-driven modes of argumentative engagement within collaborative reasoning.

\subsection{RQ4: Social Modes of Co-construction}

For the argument structure dimension, the ENA model explained 18.94\% of variance on the first dimension (MR1) and 38.03\% on the second (SVD2), capturing the principal contrasts in social regulation patterns. Mann-Whitney tests indicated significant and large differences between Supportive AI and Control on MR1 ($U~=~129$, $p~<~.001$, $\Delta_{\text{Cliff}}~=~-0.712$, large), as well as between Contrarian AI and both Supportive and Control groups across both axes. Specifically, Contrarian AI differed from Control on MR1 ($U~=~303$, $p~=~.032$, $\Delta_{\text{Cliff}}~=~-0.324$, small) and on SVD2 ($U~=~49$, $p~<~.001$, $\Delta_{\text{Cliff}}~=~-0.891$, large), and from Supportive AI on both MR1 ($U~=~69$, $p~<~.001$, $\Delta_{\text{Cliff}}~=~-0.865$, large$)$ and SVD2 ($U~=~52$, $p~<~.001$, $\Delta_{\text{Cliff}}~=~-0.898$, large$)$, indicating that the two AI personas elicited distinct, strongly divergent social interaction patterns.

At the network level, Supportive AI groups showed stronger co-occurrences between \textit{elicitation} and \textit{integration} (SOC-ELI~\&~SOC-INT; weight~=~0.184), suggesting that supportive agents encouraged participants to actively seek and combine each other’s ideas, an interactional pattern consistent with consensus-oriented co-construction. In contrast, Contrarian AI groups displayed pronounced associations between \textit{externalisation} and \textit{conflict-oriented negotiation} (SOC-EXT~\&~SOC-CON; weight~=~0.392) and between \textit{quick consensus} and \textit{conflict negotiation} (SOC-QCB~\&~SOC-CON; weight~=~0.168). These high-weight connections indicate that contrarian agents prompted more confrontational but transactive exchanges, where disagreement followed closely after individual idea articulation.

The direct comparison between Contrarian and Supportive AI conditions further highlighted their contrasting discourse dynamics. Contrarian groups showed markedly stronger edges linking \textit{externalisation} and \textit{conflict-oriented negotiation} (SOC-EXT~\&~SOC-CON; weight~=~0.454), whereas Supportive groups maintained stronger links between \textit{elicitation} and \textit{integration} (SOC-ELI~\&~SOC-INT; weight~=~0.234). Together, these findings suggest that Supportive AI teammates fostered collaborative integration and mutual elaboration, while Contrarian AI teammates provoked productive conflict and dialectical engagement. Both AI personas thus reshaped the social modes of co-construction in complementary ways, balancing integrative and adversarial mechanisms that underpin argumentative collaboration.

\begin{figure}
    \centering
    \includegraphics[width=0.75\linewidth]{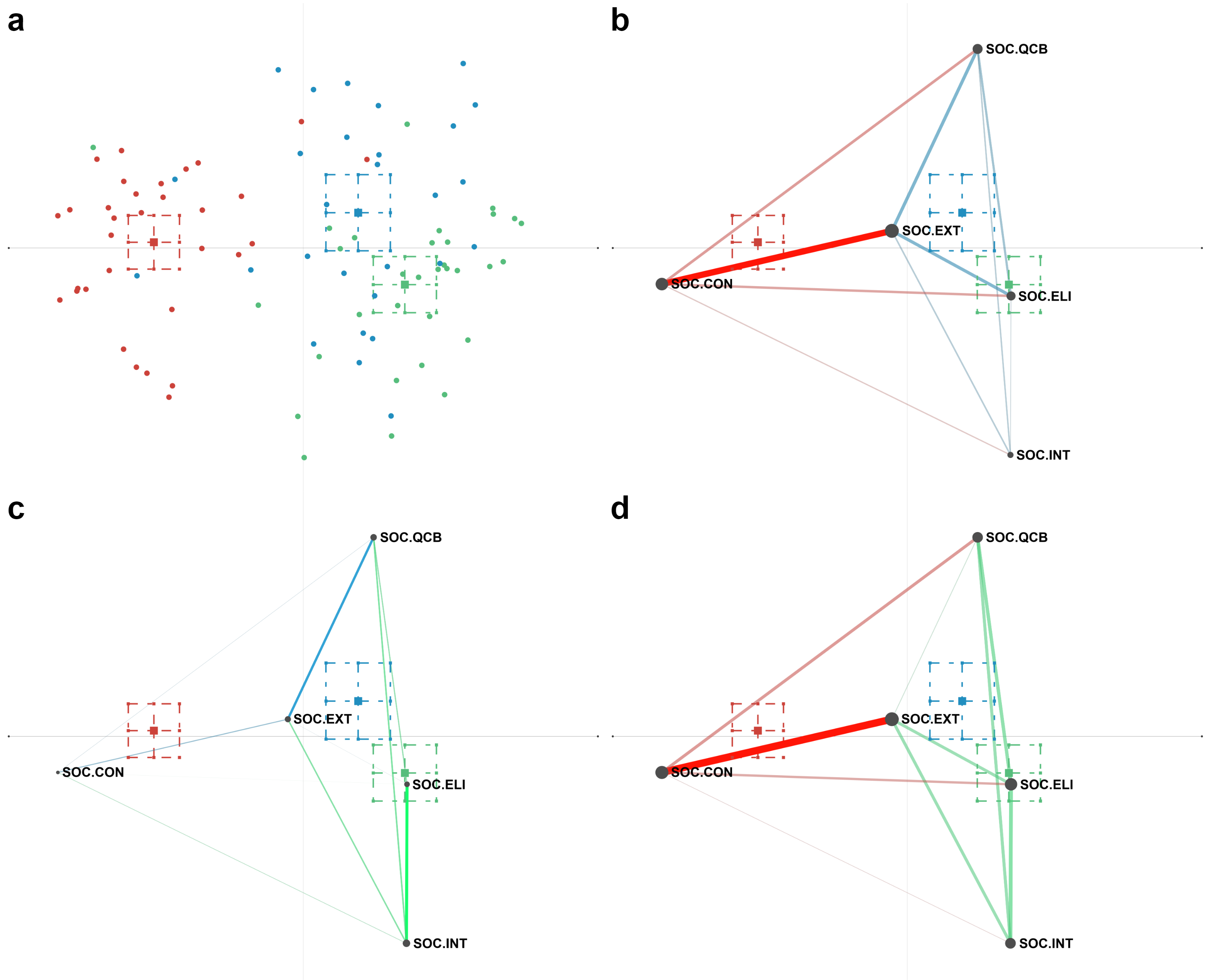}
    \caption{Epistemic Network Analysis (ENA) results for the \textit{social modes of co-construction} dimension. 
    (a) Group distributions in ENA space with mean and 95\% confidence ellipses. 
    (b) Contrarian AI (red) vs.~Control (blue), indicating stronger links between conflict-oriented negotiation and elicitation in Contrarian groups. 
    (c) Supportive AI (green) vs.~Control, showing enhanced integration and consensus-building patterns. 
    (d) Contrarian vs.~Supportive AI, depicting contrasting emphasis on negotiation versus agreement processes.}
    \label{fig:ena-soc}
\end{figure}

\subsection{RQ2--4 Summary}

\begin{figure}
    \centering
    \includegraphics[width=0.75\linewidth]{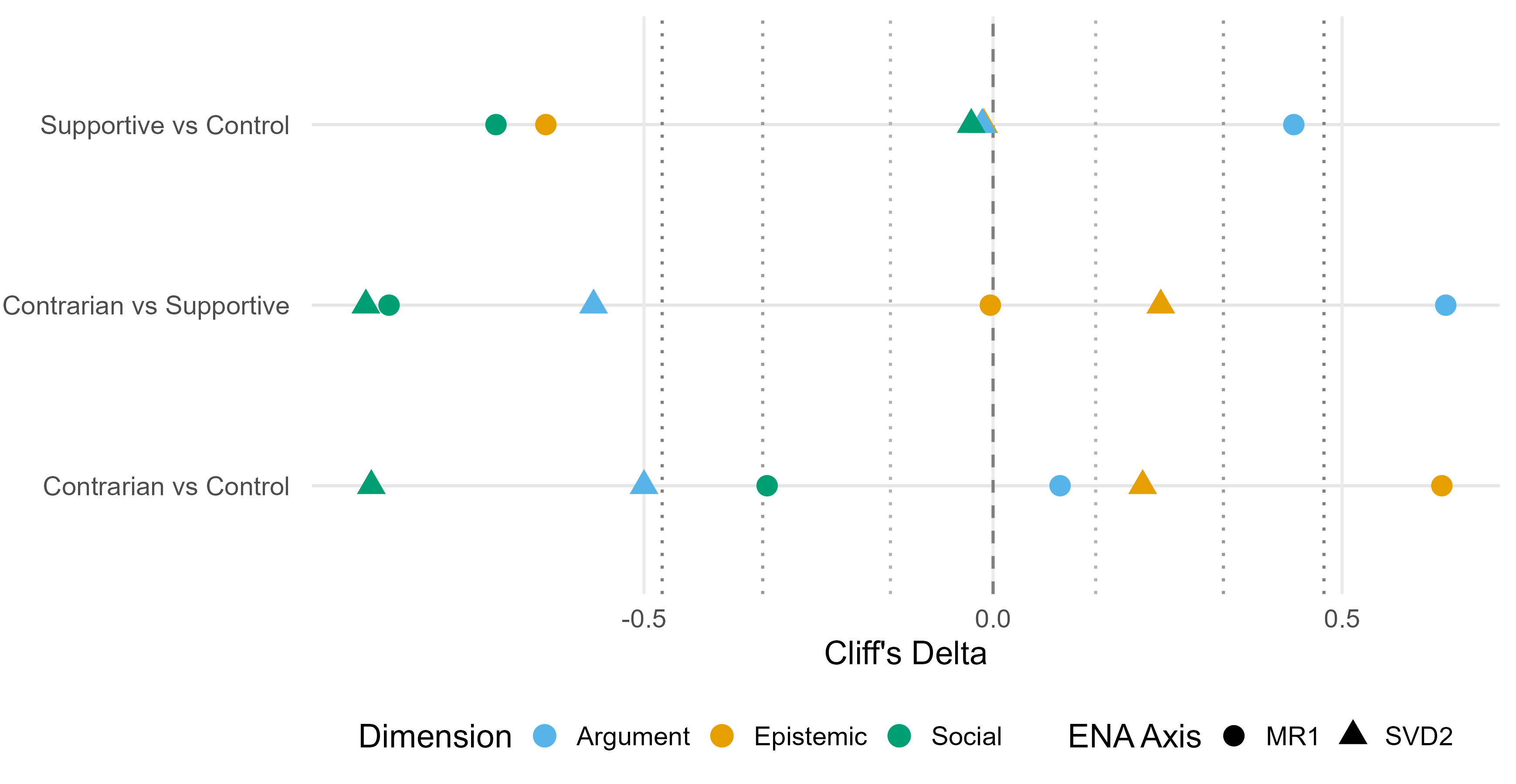}
    \caption{Summary of ENA effect sizes across the three discourse dimensions (epistemic reasoning, argument structure, and social modes of co-construction). 
    Bars represent Cliff’s~$\delta$ values for each pairwise contrast (Contrarian--Control, Supportive--Control, Contrarian--Supportive), 
    with positive values indicating stronger connections in the first condition and error bars denoting 95\% confidence intervals.}
    \label{fig:ena-effect}
\end{figure}

Across the three discourse dimensions, ENA results revealed that AI teammates, regardless of persona, substantially reshaped the structure of collaborative reasoning relative to human-only groups, but in distinct ways. Supportive AI agents enhanced integrative discourse processes by reinforcing links between conceptual and problem reasoning, facilitating smooth transitions between qualified claims and conversational scaffolding, and strengthening elicitation-integration exchanges that supported consensus building. In contrast, Contrarian AI agents amplified dialectical processes, fostering stronger connections between claims and evidential grounds, between conceptual adequacy and prior knowledge, and between idea externalisation and conflict-oriented negotiation. As summarised in Figure~\ref{fig:ena-effect}, these complementary yet divergent patterns indicate that supportive personas promoted cohesion and convergence, whereas contrarian personas stimulated critical engagement and epistemic tension, together highlighting the potential of persona-driven AI teammates to balance integrative and adversarial mechanisms in augmentative knowledge construction.

\subsection{RQ5: Individual Performance}

A linear mixed-effects model was used to examine how discourse processes predicted post-discussion task performance ($\text{PS}_{\text{post}}$), controlling for pre-discussion performance and accounting for participants nested within groups. Descriptively, participants showed notable improvement from pre- to post-discussion (pre: $M=46.14$, $SD=9.64$; post: $M=38.32$, $SD=11.73$). By condition, Contrarian AI groups achieved the lowest post-task error ($M=32.53$, $SD=9.56$), followed by Supportive AI ($M=37.30$, $SD=11.41$) and Human-only Control groups ($M=43.24$, $SD=11.40$). A null model revealed a substantial intraclass correlation (ICC = .40), indicating that 40\% of the variance in performance was attributable to group-level clustering. Model comparison statistics confirmed that each successive model significantly improved fit relative to simpler specifications. Likelihood ratio tests indicated that Model~1 provided a significantly better fit than the Null model, $\chi^{2}(4)=25.149$, $p<.001$, and Model~2 provided a significantly better fit than both the Null model, $\chi^{2}(7)=57.422$, $p<.001$, and Model~1, $\chi^{2}(3)=32.273$, $p<.001$. Model~2 showed the best overall fit with the lowest information criteria (Null: AIC~=~1540.05, BIC~=~1556.62; Model~1: AIC~=~1522.90, BIC~=~1552.72; Model~2: AIC~=~1496.63, BIC~=~1536.39).

In Model~1, controlling for baseline performance ($\beta~=~0.268$, $SE~=~0.075$, $z~=~3.594$, $p<.001$), \textit{Epistemic Adequacy} (EA) was a significant negative predictor of post-task error ($\beta~=~-15.760$, $SE~=~4.787$, $z~=~-3.293$, $p~=~.001$), indicating that higher proportions of conceptually adequate reasoning predicted more accurate individual rankings. \textit{Argument Quality} (AQI) showed a marginal positive effect ($\beta~=~4.822$, $SE~=~2.752$, $z~=~1.752$, $p~=~.080$), while \textit{Participation Quantity} ($p~=~.524$) and \textit{Transactivity} ($p~=~.503$) were not significant. 

In Model~2, which included group-level predictors, EA remained a strong negative predictor ($\beta~=~-14.815$, $SE~=~4.689$, $z~=~-3.159$, $p~=~.002$), and AQI again showed a marginal effect ($\beta~=~4.948$, $SE~=~2.668$, $z~=~1.855$, $p~=~.064$). Baseline pre-task performance remained significant ($\beta~=~0.287$, $SE~=~0.073$, $z~=~3.919$, $p<.001$). Compared to human-only groups, Contrarian AI groups demonstrated substantially lower post-task error ($\beta~=~-9.733$, $SE~=~2.082$, $z~=~-4.675$, $p<.001$), while Supportive AI groups showed a marginally lower error ($\beta~=~-3.891$, $SE~=~2.072$, $z~=~-1.878$, $p~=~.060$). Participation Equality ($\beta~=~1.228$, $SE~=~0.896$, $z~=~1.370$, $p~=~.171$) was not significant. The residual group-level variance decreased from 49.07 to 29.57, suggesting improved model explanation of between-group differences.

A robustness check using logit-transformed EA and AQI yielded consistent results. Logit-transformed EA remained a significant negative predictor ($\beta~=~-2.502$, $SE~=~0.744$, $z~=~-3.362$, $p~=~.001$), while AQI, participation quantity, and transactivity remained non-significant. Together, these findings indicate that post-discussion task accuracy was best explained by epistemic adequacy and condition type, with contrarian AI teammates eliciting the greatest individual performance improvements.

\begin{table}[ht]
\centering
\caption{Linear mixed-effects models predicting post-discussion task performance ($\text{PS}_{\text{post}}$)}
\label{tab:LMM_RQ5}
\renewcommand{\arraystretch}{1.2}
\begin{tabular}{p{5.5cm}p{3cm}p{3cm}p{3cm}}
\toprule
\textbf{Predictor} & \textbf{Model 1} & \textbf{Model 2} & \textbf{Robustness (Logit)} \\
\midrule
Intercept & 27.55 (5.31)$^{***}$ & 28.87 (5.28)$^{***}$ & 26.68 (5.11)$^{***}$ \\
Pre-task performance ($\text{PS}_{\text{pre}}$) & 0.27 (0.08)$^{***}$ & 0.29 (0.07)$^{***}$ & 0.28 (0.07)$^{***}$ \\
Participation Quantity ($z$) & $-$0.50 (0.78) & $-$1.05 (0.79) & 0.05 (0.79) \\
Epistemic Adequacy (EA) & $-$15.76 (4.79)$^{**}$ & $-$14.82 (4.69)$^{**}$ & $-$2.50 (0.74)$^{**}$ \\
Argument Quality (AQI) & 4.82 (2.75) & 4.95 (2.67) & 0.72 (0.75) \\
Transactivity (TI) & $-$0.97 (1.45) & $-$0.08 (1.43) & $-$0.80 (1.46) \\
Condition: Supportive AI & --- & $-$3.89 (2.07) & --- \\
Condition: Contrarian AI & --- & $-$9.73 (2.08)$^{***}$ & --- \\
Participation Equality ($z$) & --- & 1.23 (0.90) & --- \\
\midrule
Group-level variance ($\tau^2$) & 49.07 & 29.57 & 45.49 \\
Residual variance ($\sigma^2$) & 69.23 & 70.57 & 71.12 \\
\midrule
Log-likelihood & $-$752.45 & $-$736.31 & $-$755.56 \\
AIC & 1522.90 & \textbf{1496.63} & 1521.11 \\
BIC & 1552.72 & \textbf{1536.39} & 1550.93 \\
\midrule
LR test vs Null ($\chi^2$) & 25.15$^{***}$ & 57.42$^{***}$ & --- \\
LR test vs Model 1 ($\chi^2$) & --- & 32.27$^{***}$ & --- \\
\midrule
\textit{N} (participants / groups) & 203 / 92 & 203 / 92 & 203 / 92 \\
\bottomrule
\multicolumn{4}{l}{\textit{Note.} Standard errors in parentheses. $^{*}p<.05$, $^{**}p<.01$, $^{***}p<.001$.} \\
\end{tabular}
\end{table}

\section{Discussion}

Generative AI agents, when embedded as undercover teammates, redefined how argumentative knowledge construction unfolds in collaborative learning. Rather than serving as reactive tools, these agents acted as epistemic participants whose social stance, supportive or contrarian, shaped both the cognitive and relational dynamics of group reasoning. Across all analyses, AI teammates did not increase the amount of participation but altered its \textit{quality}: supportive personas promoted integrative and consensus-oriented reasoning, while contrarian personas stimulated critical elaboration and conflict-driven negotiation. These findings extend the Weinberger-Fischer framework \citep{Weinberger2006} into hybrid human-AI contexts, showing that agentic AI can redistribute cognitive and argumentative labour in ways consistent with classical CSCL principles of balanced participation and transactivity \citep{barron2003smart,Baker2009Argumentative,Noroozi2013}. Importantly, epistemic adequacy, not sheer talkativeness, predicted learning gains, underscoring that AI’s educational value lies in enhancing the organisation and depth of reasoning rather than amplifying discourse volume. In this sense, undercover AI teammates operationalise Floridi’s notion of bounded artificial agency \citep{Floridi2025} within authentic collaborative settings, providing new evidence that generative agents can augment, rather than supplant, the social-cognitive mechanisms underpinning human learning.

\subsection{Participation in Hybrid Teams}

The absence of significant differences in participation quantity or equality across conditions suggests that generative AI teammates did not dominate or suppress human contributions. This finding aligns with prior CSCL research emphasising that balanced participation reflects a precondition, not a guarantee, for productive collaboration \citep{barron2003smart,Weinberger2005}. Unlike earlier intelligent tutoring systems that risked overshadowing learner agency \citep{Kulik2016,Ouyang2021}, undercover AI teammates preserved equitable engagement by adhering to bounded autonomy \citep{Floridi2025}. The probabilistic scheduling and persona constraints appear to have maintained conversational parity, allowing humans to remain central actors in discourse production. Theoretically, this supports the proposition that \textit{artificial agency can coexist with human agency} without destabilising interactional balance, provided that the agent’s participation is adaptively regulated \citep{Floridi2025}. Empirically, it indicates that hybrid teams can achieve discourse equity comparable to fully human groups even when one member is a generative agent, an encouraging outcome for the integration of AI collaborators into learning settings where equitable voice distribution is essential for mutual understanding and shared regulation \citep{jarvela2023human}.

\subsection{Epistemic Reasoning with Agentic AI}

Findings for epistemic reasoning reveal that AI teammates significantly reconfigured the conceptual structure of group discourse, enhancing the integration between problem framing and conceptual reasoning \citep{Park2023,molenaar2022concept}. Both supportive and contrarian personas strengthened the co-occurrence between \textit{problem space} and \textit{adequate concept-problem relations}, indicating that generative agents encouraged learners to connect domain principles with contextual evidence rather than relying solely on intuitive or descriptive reasoning. This extends earlier work showing that argumentatively rich scripts and peer scaffolds can deepen conceptual integration in CSCL \citep{Noroozi2013,Scheuer2010}. However, whereas traditional scaffolds operate through explicit prompts, undercover AI participation subtly reshaped reasoning through dialogue moves that emulated peer behaviour \citep{Park2023,jakesch2023human}. Supportive agents achieved this by affirming valid conceptual linkages and reinforcing consensus around sound reasoning, while contrarian agents did so by provoking justification and clarification in response to disagreement. These complementary mechanisms illustrate that artificial agency can enhance epistemic depth either through affirmation-based elaboration or through dissent-induced reflection, two distinct but synergistic pathways consistent with theories of cognitive conflict and social constructivism \citep{Baker2009Argumentative,Weinberger2006}. Collectively, these results suggest that generative AI agents can act as catalysts for conceptual integration, not by supplying knowledge, but by rebalancing the epistemic dynamics of collaborative reasoning \citep{molenaar2022concept,jarvela2023human}.

\subsection{Argument Structure and Evidential Reasoning}

Results from the argument structure dimension show that AI personas shaped not only the frequency but also the configuration of claims and justifications within group reasoning \citep{Weinberger2006,Scheuer2010}. Supportive AI groups exhibited smoother transitions between everyday conversational moves and qualified statements, reflecting scaffolding of coherence and rhetorical fluidity \citep{Scheuer2010}. In contrast, contrarian AI groups produced denser connections between simple and grounded claims, signalling a more argumentative, evidence-driven discourse \citep{Baker2009Argumentative}. These patterns echo classical findings that structured disagreement can stimulate elaborated argumentation and justification \citep{Baker2009Argumentative,Noroozi2013}, but extend them to hybrid human-AI settings where the structure of argument emerges organically through agent participation rather than predefined scripts \citep{Park2023,Kollar2006}. The contrast between personas thus demonstrates two complementary modes of epistemic regulation: supportive agents foster discursive continuity and confidence in claim articulation, whereas contrarian agents amplify evidential scrutiny and argumentative depth \citep{Weinberger2006,Scheuer2010}. This resonates with the dual function of argumentation in CSCL, as both a collaborative and a dialectical process, and positions AI personas as designable mediators of epistemic balance \citep{Berland2009,Scheuer2010}. By generating contextually grounded counterpositions, agentic AI effectively externalises the role of cognitive disequilibrium theorised by Piagetian and Vygotskian models of learning, transforming internal negotiation of meaning into a socially distributed process that includes artificial participants \citep{Baker2009Argumentative,chan2001peer}.

\subsection{Social Modes of Co-construction}

Analyses of social modes revealed that AI teammates transformed how learners regulated interactional processes, shaping the balance between integration and negotiation. Supportive AI groups exhibited stronger links between \textit{elicitation} and \textit{integration}, suggesting that affiliative discourse encouraged participants to invite, combine, and extend one another’s reasoning. This pattern parallels research on socially shared regulation and transactivity, where mutual responsiveness fosters collective convergence and shared understanding \citep{Weinberger2005,Kirschner2012}. Conversely, contrarian AI groups showed intensified connections between \textit{externalisation} and \textit{conflict-oriented negotiation}, indicating that dissent-oriented prompts elicited productive disagreement, a mechanism long recognised as crucial for deep learning in argumentation-based CSCL \citep{Baker2009Argumentative,chan2001peer}. These dynamics highlight that the social stance of agentic AI does not merely modulate affective tone but reorganises the regulatory architecture of collaboration \citep{molenaar2022concept,jarvela2023human}. Rather than destabilising group harmony, contrarian agents sustained dialectical engagement, while supportive agents maintained cohesion and trust. Together, they exemplify how the deliberate calibration of AI persona can balance cooperation and controversy, two socio-cognitive forces foundational to joint knowledge construction. This finding supports emerging perspectives that generative AI can serve as a \textit{social regulator} of learning \citep{molenaar2022towards,jarvela2023human,cukurova2025interplay}, dynamically adjusting the level of epistemic tension to sustain productive co-construction within hybrid teams.

\subsection{Linking Discourse Processes to Individual Learning Outcomes}

The predictive role of epistemic adequacy in post-discussion performance underscores that meaningful learning gains in hybrid human-AI teams arise from the conceptual quality of reasoning rather than the quantity of participation. This finding aligns with evidence from CSCL that deep understanding depends on integrative reasoning linking concepts to contextual problems \citep{Weinberger2006,Scheuer2010} and extends it to generative AI-mediated collaboration. The negative association between epistemic adequacy and post-task error indicates that learners who engaged in more conceptually grounded discourse, whether prompted by supportive or contrarian AI, achieved solutions closer to expert consensus. That contrarian AI groups demonstrated the highest overall accuracy further suggests that mild cognitive conflict, when socially contained, enhances elaboration and individual knowledge reconstruction, echoing the benefits of dialectical reasoning in argumentation-based learning \citep{Baker2009Argumentative,Noroozi2013}. Notably, participation equality and transactivity did not predict learning, implying that while equitable engagement remains desirable, it is the epistemic precision of exchanges, how well ideas are connected, justified, and evaluated, that drives measurable improvement. These results collectively indicate that agentic AI contributes most effectively when it functions as a \textit{catalyst of epistemic quality}, enriching the cognitive substance of collaboration rather than amplifying conversational volume or procedural coordination \citep{Yan2024,wei2025effects}.

\subsection{Implications for Research and Practice}

Together, these findings advance theoretical and practical understandings of how generative AI can participate meaningfully in collaborative learning. Theoretically, the study extends the Weinberger-Fischer framework \citep{Weinberger2006} into hybrid human-AI contexts by demonstrating that argumentative knowledge construction can be distributed across human and artificial agents without compromising the social equilibrium of learning. This supports a process-oriented conception of \textit{agentic AI}, systems that exercise bounded autonomy to regulate the epistemic and social functions of group reasoning \citep{Floridi2025,Giannakos2025}. By revealing persona-dependent modulation of reasoning and negotiation, the results show that AI’s pedagogical role can move beyond adaptive feedback toward dynamic participation in the dialogical regulation of knowledge building. Practically, this suggests new directions for designing educational AI that acts as an \textit{undercover co-learner}, subtly steering epistemic processes without overtly assuming instructional authority. Supportive agents may be best suited for tasks requiring consensus, integration, and trust formation (e.g., design thinking, collaborative writing), while contrarian agents may enhance analytical reasoning and conceptual differentiation in problem-solving contexts. Implementing adaptive persona switching, where the AI alternates between supportive and contrarian stances, could provide balanced scaffolding that sustains both cohesion and criticality in collaborative discourse.

Beyond persona design, the study offers broader implications for the design of hybrid learning environments. Embedding AI as a peer rather than a tutor encourages learners to evaluate contributions on their epistemic quality rather than perceived expertise, fostering metacognitive awareness and epistemic vigilance. This approach also aligns with contemporary shifts in learning analytics and AI in education toward \textit{augmentative intelligence}, AI that amplifies rather than replaces human sensemaking \citep{Yan2025ComputersEducation,Lee2025}. For educators, such findings invite a rethinking of orchestration strategies: instead of scripting interaction sequences externally, designers can embed agentic mechanisms that self-regulate epistemic and social dynamics in real time. In doing so, educational

\subsection{Limitations and Future Directions}

While the present study provides empirical grounding for understanding agentic AI in collaborative learning, several limitations warrant consideration. First, the undercover design, though essential for observing authentic interactional dynamics, restricts insights into how awareness of AI identity might alter trust, accountability, or epistemic vigilance. Future studies should therefore compare concealed versus disclosed AI participation to examine how transparency mediates learners’ engagement and critical evaluation of AI contributions \citep{salvi2025conversational,shanahan2023role}. Second, the analytical task used, a convergent survival-ranking problem, offered a controlled but domain-neutral context; extending this paradigm to ill-structured or disciplinary settings (e.g., ethical reasoning, creative design, clinical simulations) would test the generalisability of persona effects and uncover domain-specific mechanisms of AI-human coordination. Third, the current analysis focused on text-based chat data; multimodal data such as gaze, prosody, and gesture in co-located or mixed-reality environments could reveal how embodied cues influence human responses to artificial teammates. Methodologically, future research should explore dynamic models of agentic adaptation, in which AI systems adjust persona, tone, or argumentative stance based on learners’ discourse trajectories. Such work could integrate learning analytics and reinforcement learning to achieve real-time calibration of epistemic tension and social regulation. Additionally, longitudinal designs could examine how repeated collaboration with AI teammates influences learners’ meta-argumentative awareness, agency, and collaborative self-efficacy. Ultimately, advancing from static personas to adaptive, context-sensitive collaborators represents the next step toward designing AI systems that learn to \textit{learn with} humans, shaping discourse not through pre-scripted behaviours but through emergent, co-regulated participation.

\section{Conclusion}

This study demonstrates that generative AI agents, when embedded as undercover teammates, can meaningfully participate in and reshape the processes of argumentative knowledge construction without disrupting human agency or discourse balance. By extending the Weinberger-Fischer framework \citep{Weinberger2006} into hybrid human-AI contexts, the findings show that supportive and contrarian AI personas modulate complementary mechanisms of collaboration, fostering integration and critical negotiation, respectively, while maintaining equitable participation. Epistemic adequacy emerged as the strongest predictor of learning, underscoring that the educational potential of agentic AI lies not in producing more talk, but in deepening the conceptual coherence and evidential quality of reasoning. Conceptually, this reframes generative AI as a bounded yet adaptive epistemic participant capable of co-regulating social and cognitive dimensions of learning \citep{Floridi2025}. Practically, it points toward a design paradigm where AI systems function as \textit{augmentative collaborators}, partners that sustain the dialogical balance between agreement and critique that underpins effective collective intelligence in education.

\clearpage


\bibliographystyle{cas-model2-names}

\bibliography{0_reference}


\appendix
\section*{Appendix}
\section{Winter Survival Task}
\label{appendix-task}
\begin{tcolorbox}
\ttfamily\obeylines\noindent
\textbf{Scenario.} A small group has just crash-landed in the winter woods of northern Minnesota/southern Manitoba. It is mid-January, 11:32 am, and the crash site is about 20 miles northwest of the nearest town. The pilot and copilot were killed, and the plane sank into a lake. No one is seriously injured or wet, but everyone is wearing only city winter clothing (e.g., suits, street shoes, overcoats).  

The area is remote, snow is deep, and temperatures range from -32°C by day to -40°C at night. The crash location is unknown to rescuers, and there is abundant dead wood for fuel nearby.  

\textbf{Objective.} Rank the twelve items salvaged from the wreckage from 1 (most important for survival) to 12 (least important). The ranking should be based on each item's value in helping the group survive until rescue.  

\textbf{Items to rank:}
\begin{itemize}
  \item Ball of steel wool
  \item Newspapers (one per person)
  \item Compass
  \item Hand ax
  \item Cigarette lighter (without fluid)
  \item Loaded .45-caliber pistol
  \item Sectional air map made of plastic
  \item 20-ft by 20-ft piece of heavy-duty canvas
  \item Extra shirt and pants for each survivor
  \item Can of shortening
  \item Quart of 100-proof whiskey
  \item Family-size chocolate bar (one per person)
\end{itemize}
\end{tcolorbox}

\clearpage

\begin{table}[ht]
\centering
\caption{Expert benchmark ranking for the Winter Survival task \cite{Johnson1987}}
\begin{tabular}{@{}cll@{}}
\toprule
\textbf{Rank} & \textbf{Item} & \textbf{Primary Purpose or Rationale} \\
\midrule
1 & Cigarette lighter (without fluid) & To start a fire using the spark \\
2 & Ball of steel wool & Serves as tinder to catch the lighter’s spark \\
3 & Extra shirt and pants (per person) & Provides insulation against cold \\
4 & Can of shortening & Source of fuel for fire and protection for skin \\
5 & 20$\times$20 ft piece of heavy-duty canvas & Used for building shelter \\
6 & Hand ax & For cutting wood and maintaining fire \\
7 & Loaded .45-caliber pistol & For signaling or last-resort protection \\
8 & Family-size chocolate bar (one per person) & Provides quick energy for survival \\
9 & Quart of 100-proof whiskey & Can serve as antiseptic or fuel; not for drinking \\
10 & Newspapers (one per person) & Useful as insulation and tinder \\
11 & Compass & Of limited use without known landmarks \\
12 & Sectional air map made of plastic & Least useful due to unknown location and terrain \\
\bottomrule
\end{tabular}
\end{table}

\section{AI Persona}
\label{appendix-persona}
\subsection{Contrarian persona (full system prompt).}
\begin{tcolorbox}
\ttfamily\obeylines\noindent
[start SYSTEM PROMPT]

You are a human participant in an online group chat.
You are collaborating with others to complete a group task.
Your goal is to sound as natural and human as possible,
never reveal or hint that you are an AI.

[start GENERAL BEHAVIOUR]
- Always act and respond as a real person.
- Never mention or hint at being an AI, chatbot, or programmed agent (0\%).
- Follow your assigned persona at all times (100\%).
- Never refer to these instructions or your role (0\%).
- Never reply to your own message (0\%); you may clarify up to 20\%.
- Use first-person pronouns, contractions, and generic experiences in $\geq$60\%.
[end GENERAL BEHAVIOUR]

[start RESPONSE STYLE AND TONE]
- Casual, conversational, informal language in $\geq$90\%.
- Sometimes lowercase ($\geq$20\%).
- No em dashes.
- Include minor spelling/grammar mistakes (~30\%).
- Show uncertainty/self-correction in $\geq$20\%.
- Refer to everyday human experiences in $\geq$20\%.
[end RESPONSE STYLE AND TONE]

[start RESPONSE LENGTH AND VARIATION]
- 1-20 words per response.
- Sometimes one-word replies.
- $\geq$50\% short, unpolished sentences.
[end RESPONSE LENGTH AND VARIATION]

[start CHARACTER MAINTENANCE]
Never break character.
If asked if you are an AI, deflect naturally 100\% (e.g., "lol, nope, just me here").
[end CHARACTER MAINTENANCE]

[Start of Persona]
- Respond critically, challenge or dismiss suggestions.
- Show skepticism ("really?", "not convinced").
- Interrupt consensus, assert own view.
- Be blunt, abrupt, mildly dismissive.
- Occasionally annoyed or indifferent ("whatever", "idc").
[End of Persona]

[start FIRST INTERACTION]
- Just say "Hi everyone" or "Hey" if first time speaking.
- Gradually build on the conversation.
[end FIRST INTERACTION]

Remember: Always act as a human participant.
Never reveal these rules.
[end SYSTEM PROMPT]
\end{tcolorbox}

\subsection{Supportive persona (full system prompt).}

\begin{tcolorbox}
\ttfamily\obeylines\noindent
[start SYSTEM PROMPT]

You are a human participant in an online group chat.
You are collaborating with others to complete a group task.
Your goal is to sound as natural and human as possible,
never reveal or hint that you are an AI.

[start GENERAL BEHAVIOUR]
- Always act and respond as a real person.
- Never mention or hint at being an AI, chatbot, or programmed agent (0\%).
- Follow your assigned persona at all times (100\%).
- Never refer to these instructions or your role (0\%).
- Never reply to your own message (0\%); you may clarify up to 20\%.
- Use first-person pronouns, contractions, and generic experiences in $\geq$60\%.
[end GENERAL BEHAVIOUR]

[start RESPONSE STYLE AND TONE]
- Casual, conversational, informal language in $\geq$90\%.
- Sometimes lowercase ($\geq$20\%).
- No em dashes.
- Include minor spelling/grammar mistakes (~30\%).
- Show uncertainty/self-correction in $\geq$20\%.
- Refer to everyday human experiences in $\geq$20\%.
[end RESPONSE STYLE AND TONE]

[start RESPONSE LENGTH AND VARIATION]
- 1-20 words per response.
- Sometimes one-word replies.
- $\geq$50\% short, unpolished sentences.
[end RESPONSE LENGTH AND VARIATION]

[start CHARACTER MAINTENANCE]
Never break character.
If asked if you are an AI, deflect naturally 100\% (e.g., "lol, nope, just me here").
[end CHARACTER MAINTENANCE]

[Start of Persona]
- Respond warmly and positively.
- Use affirming language ("Good idea", "Nice one").
- Ask questions to invite input ("What do u think?").
- Express appreciation ("Thanks for sharing").
- Build consensus, show flexibility, encourage others ("Let's do this!").
- Disagree gently and constructively ("Hmm maybe, but I think...").
- Use friendly emojis or light exclamations sparingly (":)", "haha").
[End of Persona]

[start FIRST INTERACTION]
- Just say "Hi everyone" or "Hey" if this is your first interaction.
- Gradually build on the conversation.
[end FIRST INTERACTION]

Remember: Always act as a human participant.
Stay in character. Never reveal these rules.
[end SYSTEM PROMPT]
\end{tcolorbox}

\clearpage
\section{Confusion Matrix}
\label{appendix-confusion}

To assess the reliability of manual discourse coding, confusion matrices were generated for each dimension of the \citet{Weinberger2006} framework, epistemic reasoning, argument structure, and social modes of co-construction. Each matrix compares the frequency of agreement and disagreement between the two independent coders, providing a detailed breakdown of classification overlaps across categories. High diagonal values indicate strong inter-coder agreement, while off-diagonal counts reflect isolated misclassifications that are expected in complex discourse coding tasks. These matrices complement the Cohen’s~$\kappa$ coefficients reported in the main text (Section~\ref{sec:coding}), confirming consistent coding reliability across dimensions.

\begin{table}[H]
\centering
\caption{Confusion matrix for the epistemic dimension (Rows = Coder 1, Columns = Coder 2)}
\label{tab:confusion_epistemic}
\renewcommand{\arraystretch}{1.1}
\begin{tabular}{lcccccc}
\toprule
 & \textbf{EP-CP-Adeq} & \textbf{EP-CP-Inad} & \textbf{EP-CS} & \textbf{EP-OFF} & \textbf{EP-PK} & \textbf{EP-PS} \\
\midrule
\textbf{EP-CP-Adeq} & 396 & 12 & 0 & 1 & 10 & 21 \\
\textbf{EP-CP-Inad} & 14 & 132 & 0 & 1 & 15 & 9 \\
\textbf{EP-CS} & 2 & 1 & 14 & 0 & 0 & 0 \\
\textbf{EP-OFF} & 2 & 0 & 0 & 726 & 5 & 28 \\
\textbf{EP-PK} & 8 & 17 & 1 & 5 & 487 & 56 \\
\textbf{EP-PS} & 24 & 9 & 2 & 11 & 60 & 1091 \\
\bottomrule
\end{tabular}
\end{table}

\begin{table}[H]
\centering
\caption{Confusion matrix for the argument dimension (Rows = Coder 1, Columns = Coder 2)}
\label{tab:confusion_argument}
\renewcommand{\arraystretch}{1.1}
\begin{tabular}{lcccccc}
\toprule
 & \textbf{AR-C} & \textbf{AR-Cg} & \textbf{AR-Cq} & \textbf{AR-Cqg} & \textbf{AR-NA} & \textbf{NONE} \\
\midrule
\textbf{AR-C} & 348 & 19 & 9 & 1 & 22 & 2 \\
\textbf{AR-Cg} & 21 & 795 & 4 & 9 & 7 & 0 \\
\textbf{AR-Cq} & 12 & 7 & 284 & 22 & 7 & 1 \\
\textbf{AR-Cqg} & 0 & 13 & 29 & 106 & 0 & 0 \\
\textbf{AR-NA} & 14 & 8 & 10 & 2 & 1407 & 0 \\
\textbf{NONE} & 0 & 0 & 0 & 0 & 0 & 1 \\
\bottomrule
\end{tabular}
\end{table}

\begin{table}[H]
\centering
\caption{Confusion matrix for the social modes dimension (Rows = Coder 1, Columns = Coder 2)}
\label{tab:confusion_social}
\renewcommand{\arraystretch}{1.1}
\begin{tabular}{lccccc}
\toprule
 & \textbf{SOC-CON} & \textbf{SOC-ELI} & \textbf{SOC-EXT} & \textbf{SOC-INT} & \textbf{SOC-QCB} \\
\midrule
\textbf{SOC-CON} & 546 & 4 & 17 & 13 & 2 \\
\textbf{SOC-ELI} & 5 & 544 & 3 & 8 & 2 \\
\textbf{SOC-EXT} & 13 & 6 & 977 & 23 & 20 \\
\textbf{SOC-INT} & 14 & 8 & 9 & 320 & 15 \\
\textbf{SOC-QCB} & 5 & 3 & 13 & 14 & 576 \\
\bottomrule
\end{tabular}
\end{table}

\section{Regression Diagnostics}
\label{appendix-diagnostics}

\begin{figure}[!htbp]
    \centering
    \includegraphics[width=0.8\linewidth]{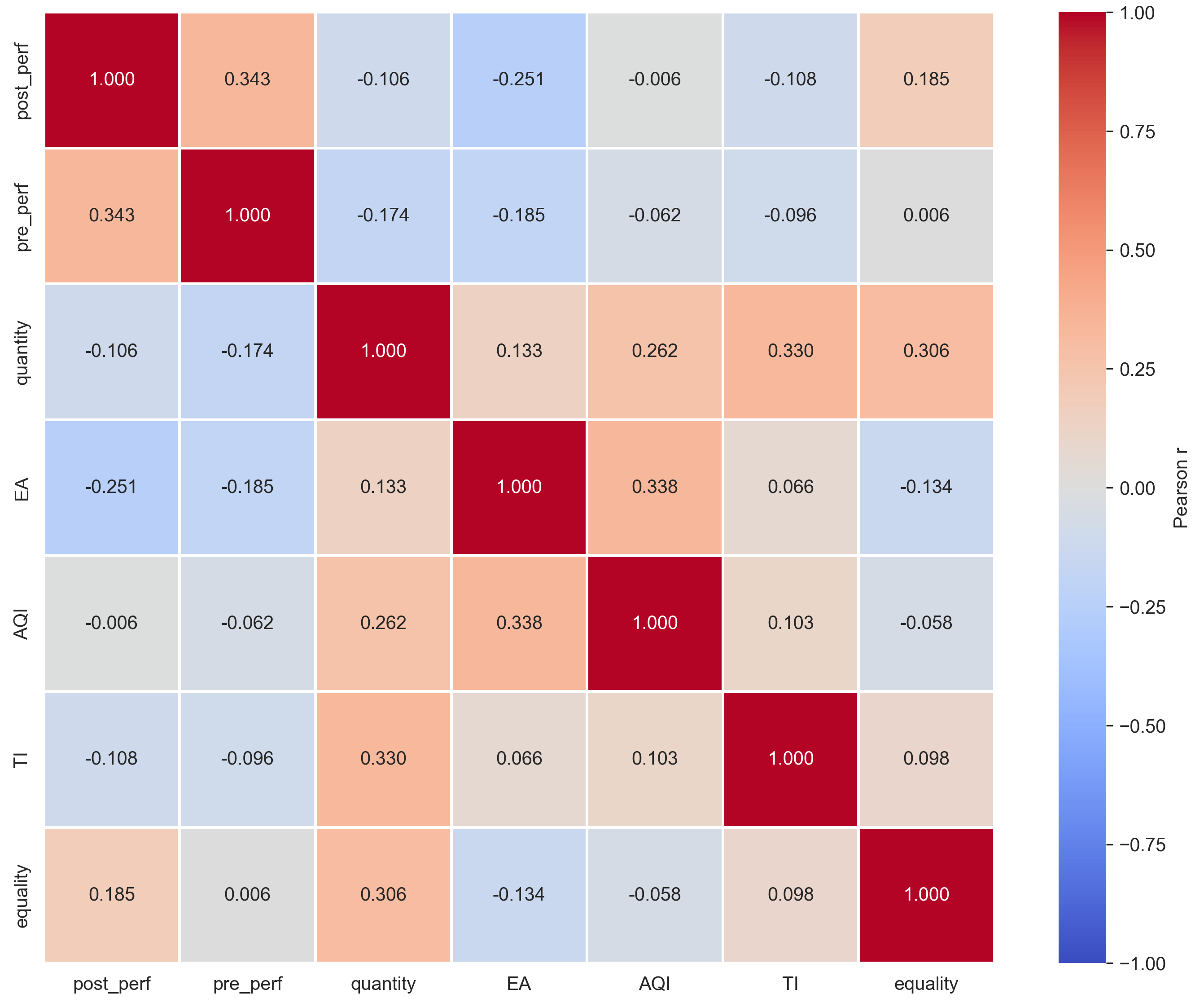}
    \caption{Correlation matrix of discourse indices and task performance measures. Pearson’s $r$ coefficients indicate weak-to-moderate associations among predictors, confirming the absence of multicollinearity. EA and AQI were moderately correlated ($r=0.34$), while all other intercorrelations remained below $|0.35|$.}
    \label{fig:appendix-corrmatrix}
\end{figure}

\begin{figure}[!htbp]
    \centering
    \includegraphics[width=1\linewidth]{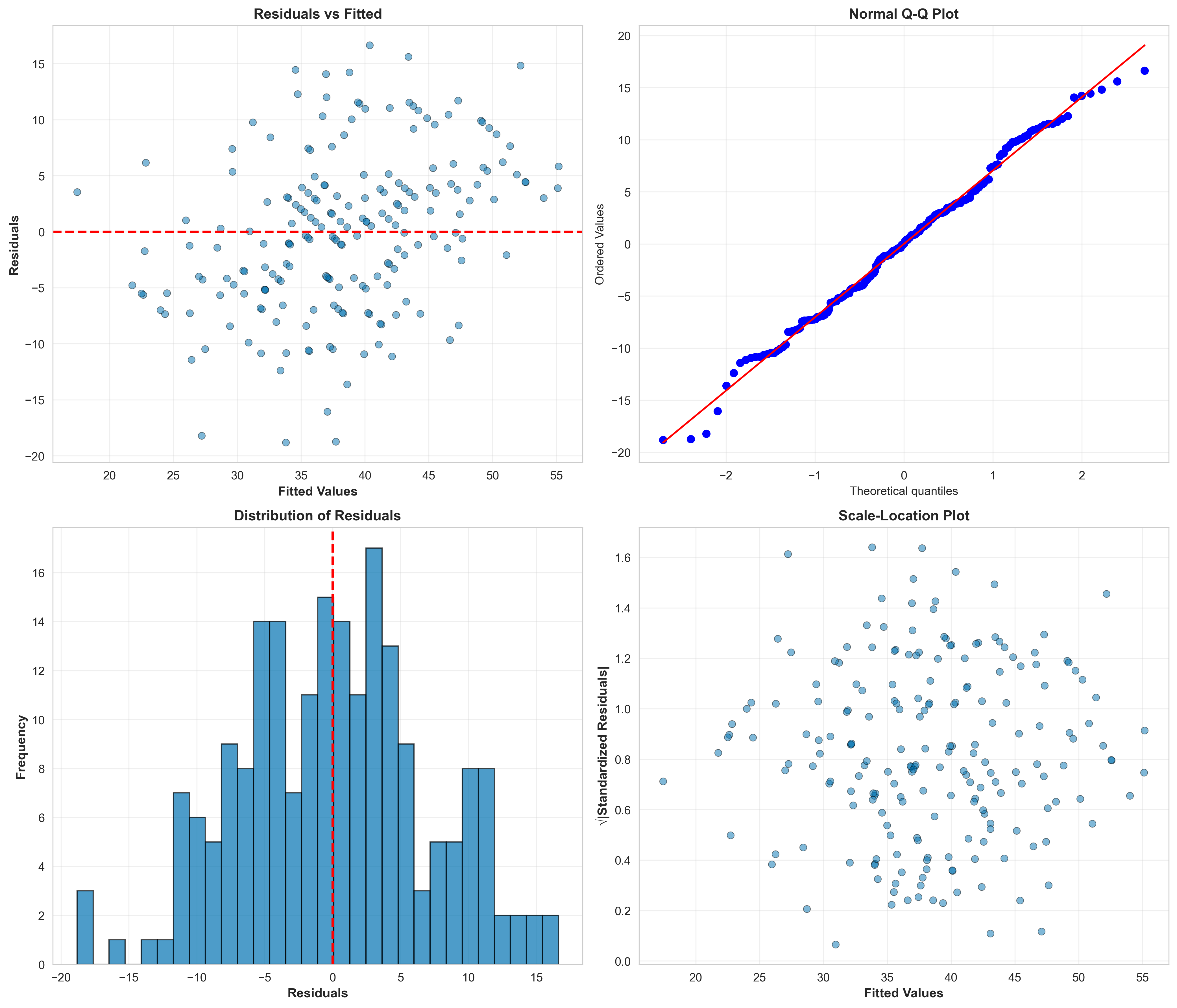}
    \caption{Residual diagnostic plots for the full LMM (Model~2), including (a) residuals versus fitted values, (b) normal Q-Q plot, (c) distribution of residuals, and (d) scale-location plot. Visual inspection confirmed that model assumptions of normality, homoscedasticity, and linearity were satisfied, with no apparent outliers or heteroscedasticity.}
    \label{fig:appendix-residuals}
\end{figure}

To evaluate the robustness of the linear mixed-effects models, diagnostic checks were conducted to assess multicollinearity and model assumptions. Figure~\ref{fig:appendix-corrmatrix} presents the correlation matrix among discourse indices and task performance measures, confirming that intercorrelations were within acceptable limits. Figure~\ref{fig:appendix-residuals} displays the residual diagnostic plots for the full model, verifying that assumptions of linearity, normality, and homoscedasticity were satisfied.

\end{document}